\documentclass[11pt,journal,draftcls,onecolumn,peerreviewca]{IEEEtran}
\usepackage{amsfonts}
\usepackage{amsmath}
\usepackage{amssymb}
\usepackage{bm}
\usepackage{xcolor,graphicx,float}
\usepackage{color, soul}
\usepackage{stfloats}
\usepackage[numbers,sort&compress]{natbib}
\usepackage[amsmath,thmmarks]{ntheorem}
\usepackage{theorem}
\usepackage{algorithm}
\usepackage{algorithmic}
\usepackage{graphicx}
\usepackage{subfigure}
\usepackage{pict2e}

\theoremheaderfont{\sc}\theorembodyfont{\upshape}
\theoremstyle{nonumberplain}
\theoremseparator{}
\theoremsymbol{\rule{1ex}{1ex}}

\begin{document}
\title{Variational Bayesian Line Spectral Estimation with Multiple Measurement Vectors}

\author{Jiang~Zhu, Qi~Zhang, Peter Gerstoft, Mihai-Alin~Badiu and Zhiwei~Xu\thanks{Part of this work has been submitted as ICASSP 2019. This journal can be seen as an extension of \cite{Confv} (parallel processing, sequential algorithm, more simulation results). Jiang Zhu, Qi Zhang and Zhiwei Xu are with the Ocean College, Zhejiang University, Zhoushan, CHINA. Peter Gerstoft is with Electrical and Computer Engineering, University of California, San Diego, USA. Mihai~Alin~Badiu is with Department of Engineering Science, University of Oxford, UK, and with Department of Electronic Systems, Aalborg University, Denmark.}
}
\maketitle
\begin{abstract}
In this paper, the line spectral estimation (LSE) problem with multiple measurement vectors (MMVs) is studied utilizing the Bayesian methods. Motivated by the recently proposed variational line spectral estimation (VALSE) method, we develop the multisnapshot VALSE (MVALSE) for multi snapshot scenarios, which is especially important in array signal processing. The MVALSE shares the advantages of the VALSE method, such as automatically estimating the model order, noise variance, weight variance, and providing the uncertain degrees of the frequency estimates. It is shown that the MVALSE can be viewed as applying the VALSE with single measurement vector (SMV) to each snapshot, and combining the intermediate data appropriately. Furthermore, the Seq-MVALSE is developed to perform sequential estimation. Finally, numerical results are conducted to demonstrate the effectiveness of the MVALSE method, compared to the state-of-the-art methods in the MMVs setting.
\end{abstract}
\section{Introduction}
\label{sec:intro}
Line spectral estimation (LSE), i.e., recovering the parameters of a superposition of complex exponential functions is one of the classical problems in signal processing fields \cite{Stoica}, which has many applications such as channel estimation in wireless communications \cite{Bajwa, HansenCom}, direction of arrival estimation in radar systems \cite{Ottersten}, speech analysis and so on. Traditional methods for solving the LSE problem include periodogram, MUSIC, ESPRIT and maximum likelihood (ML) method \cite{Stoica, Schmidt, Roy, StociaML}. For periodogram method, it is difficult to recover the closely separated frequencies \cite{Stoica}. While for subspace methods such as MUSIC and ESPRIT which utilize the covariance matrix to estimate the frequencies, they perform well when the model order is known and the signal to noise ratio (SNR) is high. As for the ML methods, it involves maximizing the nonconvex function which has a multimodal shape with a sharp global maximum. Iterative algorithm is often proposed with accurate initialization to solve the ML problem \cite{Wax1, Fleury1}.
Given that the model order is unknown in applications, some criterions such as Akaike information criterion are adopted to estimate the model order \cite{Stoica1}.

In the past decades, sparse methods for LSE have been popular due to the development of sparse signal representation and compressed sensing theory. By discretizing the continuous frequency into a finite set of grid points, the nonlinear problem can be formulated as a linear problem. $\ell_1$ optimization \cite{Malioutov}, sparse iterative covariance-based estimation (SPICE) \cite{Stoica2, Stoica3, Stoica4}, sparse Bayesian learning \cite{Gerstoft} are main sparse methods. Compared to classical methods, the grid based methods perform better by utilizing the sparsity in the spatial domain. Due to the grid mismatch, dictionary-based approaches suffer from spectral leakage. To mitigate the drawbacks of static dictionary, gridless methods have been proposed to gradually refine the dynamic dictionary, such as iterative grid refinement, joint sparse signal and parameter estimation \cite{Malioutov, Hu}. In \cite{Mamandipoor}, a Newtonalized orthogonal matching pursuit (NOMP) method is proposed, where a Newton step and feedback are utilized to refine the frequency estimation. In addition, the NOMP algorithm is also extended to deal with the MMVs setting \cite{Lin}. Compared to the incremental step in updating the frequencies in NOMP approach, the iterative reweighted approach (IRA) \cite{Fang} estimates the frequencies in parallel.

To avoid the model mismatch issues, off-grid compressed sensing methods which work directly with continuously parameterized dictionaries have been proposed \cite{Tang, Bhaskar, Yang2, Yang1, Yang3, Chi}. For the SMV case, the atom norm based method has been proposed in the noiseless case \cite{Tang}. In \cite{Bhaskar, Yang2}, the atom soft thresholding (AST) method is proposed in the noisy case. Since AST method requires knowledge of the noise variance, the gridless SPICE (GLS) method is proposed without knowledge of noise power \cite{Yang2}. In \cite{Yang1}, an exact discretization-free method called sparse and parametric approach (SPA) is proposed for uniform and sparse linear arrays, which is based on the well-established covariance fitting criterion. In \cite{Chi}, two approaches based on atomic norm minimization and structured covariance estimation are developed in the MMV case, and the benefit of including MMV is demonstrated. To further improve the resolution of the atom norm based methods, enhanced matrix completion (EMac) \cite{EMac} and reweighted atomic-norm minimization (RAM) \cite{RAM} are proposed and the resolution capability is improved numerically. These off-grid based methods involve solving a semidefinite programming (SDP) problem \cite{boyd}, whose computation complexity is prohibitively high for large-scale problems.

A different gridless approach is based on the Bayesian framework and sparse Bayesian learning (SBL) \cite{Tipping, Wipf} is adopted, where variational inference methods \cite{Shution} or maximization of the marginalized posterior probability density function (PDF) \cite{Hansenconf} is performed. For all these approaches, only point estimates of the frequency are computed in each iteration, which is similar to the classical ML methods. Another limitation is that these methods usually overestimates the model order \cite{Shution, Badiu}. In \cite{HansenTSP}, a low complexity superfast LSE methods are proposed based on fast Toeplitz matrix inversion algorithm.
\subsection{Main Contributions and Comparisons to Related Work}

In \cite{Badiu}, an off-grid variational line spectral estimation (VALSE) algorithm is proposed, where PDFs of the frequencies are estimated, instead of retaining only the point estimates of the frequencies. This more complete Bayesian approach allows to represent and operate with the frequency uncertainty, in addition to only that of the weights. Here we rigorously develop the variational Bayesian inference method for LSE in the MMVs setting, which is especially important in array signal processing. Meanwhile, the derived MVALSE reveals close relationship to the VALSE algorithm, which is suitable for parallel processing. We study the performance of the MVALSE method with von Mises prior PDFs for the frequencies. The prior information may be given from past experience, and is particularly useful when the SNR is low or few samples are available \cite{Dave}. For sequential estimation, the output of the PDF of the frequencies from the previous observations can be employed as the prior of the frequency, and sequential MVALSE (Seq-MVALSE) is proposed. Furthermore, substantial experiments are conducted to illustrate the competitive performance of the MVALSE method and its application to DOA problems, compared to other sparse based approaches.
\subsection{Paper Organization and Notation}
The rest of this paper is organized as below. Section \ref{setup} describes the signal model with MMV and introduces the probabilistic formulation. Section \ref{Algorithm} develops the MVALSE algorithm and the details of the updating expressions are presented. In addition, the relationship between the VALSE and MVALSE are revealed, and the Seq-MVALSE is also presented. Substantial numerical experiments are provided in Section \ref{NS} and Section \ref{conclusion} concludes the paper.

Let ${\mathcal S}\subset \{1,\cdots,N\}$ be a subset of indices and $|{\mathcal S}|$ denote its cardinality. For the matrix ${\mathbf A}\in{\mathbb C}^{M\times N}$, let $\mathbf A_{\mathcal S}$ denote the submatrix by deleting the columns of $\mathbf A$ indexed by $\{1,\cdots,N\}\backslash {\mathcal S}$. For the matrix $\mathbf H\triangleq[\mathbf h_1^{\rm T};\cdots;\mathbf h_N^{\rm T}]\in{\mathbb C^{N\times L}}$ and $\mathbf W\triangleq[\mathbf w_1^{\rm T};\cdots;\mathbf w_N^{\rm T}]\in{\mathbb C^{N\times L}}$, let $\mathbf h_i^{\rm T}$ and $\mathbf w_i^{\rm T}$ denote the $i$th row of $\mathbf H$ and $\mathbf W$, respectively. Let $\mathbf H_{\mathcal S}$ and $\mathbf W_{\mathcal S}$ denote the submatrix by choosing the rows of $\mathbf H$ and $\mathbf W$ indexed by ${\mathcal S}$. For the matrix $\mathbf J\in\mathbb C^{N\times N}$, let $\mathbf J_{\mathcal S}$ denote the submatrix by choosing both the rows and columns of $\mathbf J$ indexed by $\mathcal S$. Let ${(\cdot)}^{*}_{\mathcal S}$, ${(\cdot)}^{\rm T}_{\mathcal S}$ and ${(\cdot)}^{\rm H}_{\mathcal S}$ be the conjugate, transpose and Hermitian transpose operator of ${(\cdot)}_{\mathcal S}$, respectively. Let $\mathbf I_L$ denote the identity matrix of dimension $L$. Let $||\cdot||_{\rm F}$ denote the Frobenius norm. $``\sim i"$ denotes the indices ${\mathcal S}$ excluding $i$ and ${\rm Re}\{\cdot\}$ returns the real part. Let ${\mathcal {CN}}({\mathbf x};{\boldsymbol \mu},{\boldsymbol \Sigma})$ denote the complex normal distribution of ${\mathbf x}$ with mean ${\boldsymbol \mu}$ and covariance ${\boldsymbol \Sigma}$, and let ${\mathcal {VM}}(\theta,\mu,\kappa)$ denote the von Mises distribution of $\theta$ with mean direction $\mu$ and concentration parameter $\kappa$. For a vector $\mathbf x$, let $\|{\mathbf x}\|_0$ denote the number of nonzero elements, and sometimes we let $[{\mathbf x}]_i$ or $x_i$ denote its $i$th element. Similarly, let $[{\mathbf B}]_{i,j}$ or $B_{ij}$ denote the $(i,j)$th element of ${\mathbf B}$, and let ${\mathbf B}_{i,:}$ and ${\mathbf B}_{:,j}$ denote the $i$th row and $j$th column of ${\mathbf B}$, respectively.
\section{Problem Setup}\label{setup}
 For line spectral estimation problem with $L$ snapshots, the measurements $\mathbf Y\in\mathbb{C}^{M\times L}$ consist of a superposition of $K$ complex sinusoids corrupted by the additive white Gaussian noise (AWGN) ${\mathbf U}$, which is described by
\begin{align}\label{signal-generate}
{\mathbf Y} = \sum\limits_{k=1}^{K}{\mathbf a}({\widetilde\theta}_k){\widetilde{\mathbf w}}_k^{\rm T}+{{\mathbf U}},
\end{align}
where $M$ is the number of measurements for each observation. The complex weights over the $L$ snapshots and the frequency of the $k$th component are represented by ${\widetilde{\mathbf w}}_k\in \mathbb{C}^{L\times 1}$ and respectively ${\widetilde\theta}_k \in [-\pi,\pi)$. The elements of the noise ${\mathbf U} \in {\mathbb C}^{M\times L}$ are i.i.d. and $U_{ij}\sim {\mathcal {CN}}(U_{ij};0,\nu)$, and ${\mathbf a}(\widetilde\theta_k)=\left[1,{\rm e}^{{\rm j}{\widetilde\theta}_k},\cdots,{\rm e}^{{\rm j}(M-1){\widetilde\theta}_k}\right]^{\rm T}$.

Since the number of complex sinusoids $K$ is generally unknown, the measurements ${\mathbf Y}$ is assumed to consist of a superposition of known $N$ components with $N>K$ \cite{Badiu}, i.e.,
\begin{align}\label{signal-model}
{\mathbf Y} = \sum\limits_{i=1}^{N}{\mathbf a}(\theta_i){\mathbf w}_i^{\rm T}+{\mathbf U} = {\mathbf A}{\mathbf W} + {\mathbf U},
\end{align}
where ${\mathbf A} = \left[{\mathbf a}(\theta_1),\cdots,{\mathbf a}(\theta_N)\right]\in\mathbb{C}^{M\times{N}}$, $\mathbf a(\theta_i)$ denotes the $i{\rm th}$ column of $\mathbf A$, ${\mathbf w}_i^{\rm T}$ denote the $i$th row of $\mathbf W \in\mathbb{C}^{N\times{L}}$. Since $N>K$, the binary hidden variables ${\mathbf s}=[s_1,...,s_N]^{\rm T}$ are introduced and the probability mass function is $p({\mathbf s};\rho)=\prod_{i=1}^Np(s_i;\rho)$, where $s_i\in\{0,1\}$ and
\begin{small}
\begin{align}\label{pmfs}
p(s_i;\rho) = \lambda^{s_i}(1-\lambda)^{(1-s_i)}.
\end{align}
\end{small}We assume $p({\mathbf W}|{\mathbf s};\tau)=\prod_{i=1}^Np({\mathbf w}_i|s_i;\tau)$, where $p({\mathbf w}_i|s_i;\tau)$ follows a Bernoulli-Gaussian distribution
\begin{small}
\begin{align}\label{pdfws}
p({\mathbf w_i}|s_i;\tau) = (1 - s_i){\delta}({\mathbf w_i}) + s_i{\mathcal {CN}}({\mathbf w_i};{\mathbf 0},\tau{\mathbf I}_L),
\end{align}
\end{small}where ${\delta}(\cdot)$ is the Dirac delta function. From (\ref{pmfs}) and (\ref{pdfws}), it can be seen that $\lambda$ controls the probability of the $i$th component being active. The prior distribution $p({\boldsymbol \theta})$ of the frequency ${\boldsymbol \theta} = [\theta_1,...,\theta_N]^{\rm T}$ is $p({\boldsymbol \theta}) = \begin{matrix} \prod_{i=1}^N p(\theta_i) \end{matrix}$, where $p(\theta_i)$ is encoded through the von Mises distribution \cite[p.~36]{Direc}
\begin{small}
\begin{align}\label{prior_theta}
p(\theta_i) = {\mathcal {VM}}(\theta_i;\mu_{0,i},\kappa_{0,i})= \frac{1}{2\pi{I_0}(\kappa_{0,i})}{\rm e}^{\kappa_{0,i}{\cos(\theta-\mu_{0,i})}},
\end{align}
\end{small}where $\mu_{0,i}$ and $\kappa_{0,i}$ are the mean direction and concentration parameters of the prior of the $i$th frequency $\theta_i$, $I_p(\cdot)$ is the modified Bessel function of the first kind and the order $p$ \cite[p.~348]{Direc}.
Note that $\kappa_{0,i}=0$ corresponds to the uninformative prior distribution $p(\theta_i) = {1}/({2\pi})$ \cite{Badiu}.

For measurement model
(\ref{signal-model}), the likelihood $p({\mathbf Y}|{\mathbf A}{\mathbf W};\nu)$ is
\begin{align}
p({\mathbf Y}|{\mathbf A}{\mathbf W};\nu)=\prod\limits_{i,j}{\mathcal {CN}}(Y_{ij};[{\mathbf A}{\mathbf W}]_{i,j},\nu).
\end{align}
Let ${\boldsymbol \beta} = \{\nu,~\rho,~\tau\}$ and ${\boldsymbol \Phi}=\{{\boldsymbol \theta}, {\mathbf W}, {\mathbf s}\}$ be the model and estimated parameters. Given the above statistical model, the type II maximum likelihood (ML) estimation of the model parameters $\hat{\boldsymbol \beta}_{\rm ML}$ is
\begin{align}\label{MLbeta}
\hat{\boldsymbol \beta}_{\rm ML}=\underset{{\boldsymbol \beta}}{\operatorname{argmax}}~\int p({\mathbf Y}, \boldsymbol\Phi;{\boldsymbol\beta})
{\rm d}{\mathbf s}{\rm d}{\mathbf W}{\rm d}{\boldsymbol \theta},
\end{align}
where $p({\mathbf Y}, \boldsymbol\Phi;{\boldsymbol\beta})\propto p({\mathbf Y}|{\mathbf A}{\mathbf W};\nu)\prod_{i=1}^N p(\theta_i)p({\mathbf w_i}|s_i;\tau)p(s_i;\rho)$.
Then the minimum mean square error (MMSE) estimate ${\boldsymbol \Phi}_{\rm MMSE}$ of the parameters ${\boldsymbol \Phi}$ is
\begin{align}\label{MLPhi}
&\hat{\boldsymbol \Phi}_{\rm MMSE}={\rm E}[{\boldsymbol \Phi}|{\mathbf Y};\hat{\boldsymbol \beta}_{\rm ML}],
\end{align}
where the expectation is taken with respect to the PDF
\begin{align}
p({\boldsymbol \Phi}|{\mathbf Y};\hat{\boldsymbol \beta}_{\rm ML})\propto p({\mathbf Y}|{\mathbf A}{\mathbf W};\hat{\nu}_{\rm ML})\prod_{i=1}^N p(\theta_i)p({\mathbf w_i}|s_i;\hat{\tau}_{\rm ML})p(s_i;\hat{\rho}_{\rm ML}).
\end{align}
However, computing both the ML estimate of ${\boldsymbol \beta}$ (\ref{MLbeta}) and the MMSE estimate of ${\boldsymbol \Phi}$ (\ref{MLPhi}) are intractable. Thus an iterative algorithm is designed in the following.
\section{MVALSE Algorithm}\label{Algorithm}
In this section, a mean field variational Bayes method is proposed to find an approximate PDF $q({\boldsymbol \Phi}|{\mathbf Y})$ by minimizing the Kullback-Leibler (KL) divergence ${\rm{KL}}(q({\boldsymbol \Phi}|{\mathbf Y})||p({\boldsymbol \Phi}|{\mathbf Y}))$  \cite[p.~732]{Murphy}
\begin{align}\label{KLdef}
{\rm{KL}}(q({\boldsymbol \Phi}|{\mathbf Y})||p({\boldsymbol \Phi}|{\mathbf Y}))=\int q({\boldsymbol \Phi}|{\mathbf Y})\ln \frac{q({\boldsymbol \Phi}|{\mathbf Y})}{p({\boldsymbol \Phi}|{\mathbf Y})}{\rm d}{\boldsymbol \theta}{\rm d}{\mathbf W}{\rm d}{\mathbf s}.
\end{align}
For any assumed PDF $q({\boldsymbol \Phi}|{\mathbf Y})$, the log marginal likelihood (model evidence) $\ln p({\mathbf Y;\boldsymbol\beta})$  is \cite[pp.~732-733]{Murphy}
\begin{align}\label{DL}
\ln p({\mathbf Y;\boldsymbol\beta})={\rm{KL}}(q({\boldsymbol \Phi}|{\mathbf Y})||p({\boldsymbol \Phi}|{\mathbf Y}))+ {\mathcal L}(q({\boldsymbol \Phi}|{\mathbf Y});{\boldsymbol \beta}),
\end{align}
where
\begin{align}\label{DL-L}
{\mathcal L}(q({\boldsymbol \Phi}|{\mathbf Y});{\boldsymbol \beta}) = {\rm E}_{q({\boldsymbol \Phi}|{\mathbf Y})}\left[\ln{\tfrac{p({\mathbf Y},{\boldsymbol \Phi;\boldsymbol \beta})}{q({\boldsymbol\Phi}|{\mathbf Y})}}\right].
\end{align}
For a given data $\mathbf Y$, $\ln p({\mathbf Y;\boldsymbol\beta})$ is a constant, thus minimizing the KL divergence is equivalent to maximizing ${\mathcal L}(q({\boldsymbol \Phi}|{\mathbf Y});{\boldsymbol\beta})$ in (\ref{DL}). Therefore we maximize ${\mathcal L}(q({\boldsymbol \Phi}|{\mathbf Y});{\boldsymbol\beta})$ in the sequel.

For the factored PDF $q({\boldsymbol \Phi}|{\mathbf Y})$, the following assumptions are made:
\begin{itemize}
  \item Given $\mathbf Y$, the frequencies $\{\theta_i\}_{i=1}^N$ are mutually independent.
  \item The posterior of the binary hidden variables $q({\mathbf {s|Y}})$ has all its mass at $\widehat{\mathbf s}$, i.e., $q({\mathbf s}|{\mathbf Y})=\delta({\mathbf {s-\widehat{s}}})$.
  \item Given $\mathbf Y$ and $\mathbf s$, the frequencies and weights are independent.
\end{itemize}
As a result, $q({\boldsymbol \Phi}|{\mathbf Y})$ can be factored as
\begin{align}
q({\boldsymbol \Phi}|{\mathbf Y}) = \prod_{i=1}^Nq(\theta_i|\mathbf Y)q({\mathbf {W|Y,s}})\delta({\mathbf {s-\widehat{s}}}).\label{postpdf}
\end{align}
Due to the factorization property of (\ref{postpdf}), the frequencies ${\boldsymbol \theta}$ can be estimated from the marginal distribution $q({\boldsymbol \Phi}|{\mathbf Y})$ as \cite[pp.~26]{Direc}
\begin{subequations}\label{ahat}
\begin{align}
&\widehat{\theta}_i = {\rm arg}({\rm E}_{q{(\theta_i|\mathbf Y})}[{\rm e}^{{\rm j}\theta_i}]),\label{ahat_a}\\
&\widehat{\mathbf a}_i = {\rm E}_{q{(\theta_i|\mathbf Y)}}[\mathbf a(\theta_i)],~i\in\{1,...,N\},\label{ahat_b}
\end{align}
\end{subequations}
where ${\rm arg}(\cdot)$ returns the angle. In Section \ref{yita}, $q(\theta_i|\mathbf Y)$ is approximated as a von Mises distribution. For von Mises distribution ${\mathcal {VM}}(\theta;\mu,\kappa)$ (\ref{prior_theta}), ${\rm arg}({\rm E}_{{\mathcal {VM}}(\theta;\mu,\kappa)}[{\rm e}^{{\rm j}\theta}]) = {\rm arg}\left({\rm e}^{{\rm j}\mu}\frac{I_1(\kappa)}{I_0(\kappa)}\right) = \mu = {\rm E}_{{{\mathcal {VM}}(\theta;\mu,\kappa)}}[\theta]$. Therefore, ${\widehat\theta}_i$ is also the mean direction of $\theta$ for von Mises distribution. Besides, ${\rm E}[{\rm e}^{{\rm j}m\theta}]={\rm e}^{{\rm j}m\mu}I_m(\kappa)/I_0(\kappa)$ \footnote{As $I_m(\kappa)/I_0(\kappa)<1$ for $m\in {1,\cdots,M-1}$, the magnitudes of the elements of ${\rm E}_{q{(\theta_i|\mathbf Y)}}[\mathbf a(\theta_i)]$ are less than $1$. An alternative approach is to assume the following posterior PDF $\delta(\theta_i-\widehat{\theta}_i)$ which corresponds to the point estimates of the frequencies, and let $\widehat{\mathbf a}_i$ be ${\mathbf a}(\widehat{\theta}_i)$, which yields the VALSE-pt algorithm \cite{Badiu}. Numerical results show that the performance of VALSE-pt is slightly worse than that of VALSE algorithm \cite{Badiu}. Here we use (\ref{ahat_b}) to estimate $\mathbf a(\theta_i)$.}.

Given that $q({\mathbf s|\mathbf Y}) = \delta(\mathbf {s-\widehat{s}})$, the posterior PDF of $\mathbf W$ is
\begin{align}\label{w_s}
q({\mathbf W}|{\mathbf Y}) = \int q({\mathbf W}|{\mathbf Y},{\mathbf s})\delta({\mathbf s}-\widehat{{\mathbf s}}){\rm d}{\mathbf s} = q({\mathbf W}|{\mathbf Y};\widehat{\mathbf s}).
\end{align}
For the given posterior PDF $q(\mathbf W|{\mathbf Y})$, the mean and covariance of the weights are estimated as
\begin{subequations}\label{w_est}
\begin{align}
&\widehat{\mathbf w}_i = {\rm E}_{q({\mathbf {W|Y}})}[\mathbf w_i],\\
&\widehat{\mathbf C}_{i,j} = {\rm E}_{q{(\mathbf {W|Y}})}[{\mathbf {w}_i{\mathbf w}_j^{\rm H}}] - {\widehat{\mathbf w}}_i\widehat{\mathbf w}_j^{\rm H},~ i,j\in\{1,...,N\}.
\end{align}
\end{subequations}
Let $\mathcal S$ be the set of indices of the non-zero components of $s$, i.e.,
\begin{align}\notag
\mathcal S = \{i|1\leq i\leq N,s_i = 1\}.
\end{align}
Analogously, $\widehat{\mathcal S}$ is defined based on $\widehat{\mathbf s}$. The model order is estimated as the cardinality of $\widehat{\mathcal S}$, i.e.,
\begin{align}\notag
\widehat{K} = |\widehat{\mathcal S}|.
\end{align}
According to (\ref{signal-model}), the noise-free signal is reconstructed as
\begin{align}\notag
\widehat{\mathbf X} = \sum_{i\in{\widehat{\mathcal S}}}\widehat{\mathbf a}_i\widehat{\mathbf w}_i^{\rm T}.
\end{align}

Maximizing ${\mathcal L}(q({\boldsymbol \Phi}|{\mathbf Y}))$ with respect to all the factors is also intractable. Similar to the Gauss-Seidel method \cite{Bertsekas}, $\mathcal L$ is optimized over each factor $q(\theta_i|\mathbf Y)$, $i=1,\dots,N$ and $q({\mathbf {W,s|Y}})$ separately with the others being fixed. Let ${\mathbf z}=(\theta_1,\dots,\theta_N,({\mathbf W},{\mathbf s}))$ be the set of all latent variables. Maximizing ${\mathcal L}(q({\boldsymbol \Phi}|{\mathbf Y});{\boldsymbol \beta})$ (\ref{DL-L}) with respect to the posterior approximation $q({\mathbf z}_d|{\mathbf Y})$ of each latent variable ${\mathbf z}_d,~d=1,\dots,N+1$ yields \cite[pp. 735, eq. (21.25)]{Murphy}
\begin{align}\label{upexpression}
\ln q({\mathbf z}_d|{\mathbf Y})={\rm E}_{q({{\mathbf z}\setminus{\mathbf z}_d}|{\mathbf Y})}[\ln p({\mathbf Y},{\mathbf z})]+{\rm const},
\end{align}
where the expectation is with respect to all the variables ${\mathbf z}$ except ${\mathbf z}_d$ and the constant ensures normalization of the PDF. In the following, we detail the procedures.
\subsection{Inferring the frequencies}\label{yita}
For each $i = 1,...,N$, we maximize $\mathcal L$ with respect to the factor $q({\theta_i|\mathbf Y})$. For $i\notin{\mathcal S}$, we have $q(\theta_i|\mathbf Y)= p(\theta_i)$. According to (\ref{upexpression}), for $i\in{\mathcal S}$, the optimal factor $q(\theta_i|\mathbf Y)$ can be calculated as
\begin{align}\label{qtheta_cal}
\ln q(\theta_i|\mathbf Y) = &{\rm E}_{q({\mathbf z}\setminus{\theta_i}|{\mathbf Y})}\left[
\ln p({\mathbf Y}, \boldsymbol\Phi;{\boldsymbol\beta})\right]+ \rm const.
\end{align}
In Appendix \ref{derqtheta}, it is shown that
\begin{align}\label{pdf-q}
q({\theta}_i|{\mathbf Y})\propto \underbrace{p(\theta_i)}_{(a)}\underbrace{\exp({\rm Re}\{\boldsymbol \eta_i^{\rm H}\mathbf a(\theta_i)\})}_{(b)},
\end{align}
where the complex vector $\boldsymbol\eta_i$ is given by
\begin{align}\label{yita-i}
\boldsymbol\eta_i = \frac{2}{\nu}\left({\mathbf Y}-\sum_{j\in\widehat{\mathcal S} \backslash \{i\}}\widehat{\mathbf a}_j{\widehat{\mathbf w}}_j^{\rm T}\right){\widehat{\mathbf w}}_i^*-\frac{2}{\nu}\sum_{j\in\widehat{\mathcal S} \backslash \{i\}}{\rm tr}({\widehat{\mathbf C}}_{j,i})\widehat{\mathbf a}_j
\end{align}
for $i\in \widehat{\mathcal S}$, and ${\boldsymbol\eta}_i={\mathbf 0}$ otherwise, which is consistent with the results in \cite[equ. (17)]{Badiu} for the SMV case. In order to obtain the approximate posterior distribution of $\mathbf W$, as shown in the next subsection, (\ref{ahat_b}) needs to be computed. While it is hard to obtain the analytical results for the PDF (\ref{pdf-q}), heuristic $2$ from \cite{Badiu} is used to obtain a von Mises approximation. For the second frequency, the prior can be similarly chosen from the set $\{p(\theta_i)\}_{i=1}^N$ with the first selected prior being removed. For the other frequencies, the steps follow similarly.

It is worth noting that for the prior distribution (\ref{prior_theta}), when $\kappa_p$ tends to infinity, $p(\theta_i) = \delta(\theta-\mu_{0,i})$, where $\delta(\cdot)$ denotes the Dirac delta function. Consequently, the signal model (\ref{signal-model}) is a sum over deterministic frequencies $\mu_{0,i}$, i.e., ${\mathbf Y} = \sum\limits_{i=1}^{N}{\mathbf a}(\mu_{0,i}){\mathbf w}_i^{\rm T}+{\mathbf U}$. Thus, in this case, the MVALSE algorithm is a complete grid based method. When $\kappa_p = 0$, $p(\theta_i) = \frac{1}{2\pi}$ corresponding to the uninformative prior, the VALSE is a complete off-grid based method. Thus, by varying $\kappa_p$, the prior of the VALSE algorithm provides a trade-off between grid method and off-grid method.
\subsection{Inferring the weights and support}\label{W-and-C}
Next $q(\theta_i|{\mathbf Y}),i=1,...,N$ are fixed and $\mathcal L$ is maximized w.r.t. $q({\mathbf W},{\mathbf s}|{\mathbf Y})$.
Define the matrices $\mathbf J$ and $\mathbf H$ as
\begin{subequations}\label{J-H}
\begin{align}
&{J}_{ij}= 	
\begin{cases}
M,&i=j\\
{\widehat{\mathbf a}}^{\rm H}_i{\widehat{\mathbf a}}_j,&i\neq{j}
\end{cases},\quad i,j\in\{1,2,\cdots,N\},\\
&{\mathbf H} = \widehat{\mathbf A}^{\rm H}{\mathbf Y},
\end{align}
\end{subequations}
where ${J}_{ij}$ denotes the $(i,j)$th element of $\mathbf J$.

According to (\ref{upexpression}), $q({\mathbf W},{\mathbf s}|{\mathbf Y})$ can be calculated as
\begin{align}
&\ln q({\mathbf W},{\mathbf s}|{\mathbf Y}) = {\rm E}_{q({\mathbf z}\setminus{({\mathbf W,s})}|{\mathbf Y})}\left[
\ln p({\mathbf Y},{\boldsymbol \Phi};{\boldsymbol\beta})\right]+ {\rm const}\notag\\
=& {\rm E}_{q{(\boldsymbol\theta|{\mathbf Y}})}[\sum_{i=1}^N \ln p(s_i) + \ln p(\mathbf {W|s})+\ln p(\mathbf {Y}|\boldsymbol \theta,\mathbf W)]+{\rm const}\notag\\
=&||{\mathbf s}||_0\ln\frac{\lambda}{1-\lambda} + ||{\mathbf s}||_0L\ln\frac{1}{\pi\tau}- \frac{1}{\tau}{\rm tr}(\mathbf W_{{\mathcal S}}\mathbf W^{\rm H}_{{\mathcal S}})+ \frac{2}{\nu}{\rm Re}\{{\rm tr}({\mathbf W}^{\rm H}_{{\mathcal S}}{\mathbf H}_{\mathcal S})\}-\frac{1}{\nu}{\rm tr}(\mathbf W^{\rm H}_{{\mathcal S}}\mathbf J_{{\mathcal S}}\mathbf W_{{\mathcal S}}) + {\rm const}\notag\\
=&{\rm tr}\left(({\mathbf W}_{\mathcal S} - \widehat{\mathbf W}_{{\mathcal S}})^{\rm H}\widehat{\mathbf C}_{{\mathcal S},0}^{-1}({\mathbf W}_{\mathcal S} - \widehat{\mathbf W}_{{\mathcal S}})\right) + {\rm const},\label{Ws_pos}
\end{align}
where
\begin{subequations}\label{W-C-1}
\begin{align}
&\widehat{\mathbf W}_{{\mathcal S}} = \nu^{-1}\widehat{\mathbf C}_{{\mathcal S},0}\mathbf H_{{\mathcal S}},\label{What}\\
&\widehat{\mathbf C}_{{\mathcal S},0} = \left(\frac{{\mathbf J}_{{\mathcal S}}}{\nu}+\frac{{\mathbf I}_{|{\mathcal S}|}}{\tau}\right)^{-1}.\label{defCshat0}
\end{align}
\end{subequations}

From (\ref{postpdf}), the posterior approximation $q({\mathbf W},{\mathbf s}|{\mathbf Y})$  can be factored as the product of $q({\mathbf W}|{\mathbf Y},{\mathbf s})$ and $\delta({\mathbf s}-{\widehat{\mathbf s}})$. According to the formulation of (\ref{Ws_pos}), for a given $\widehat{\mathbf s}$, $q({\mathbf W}|{\mathbf Y})$ is a complex Gaussian distribution, i.e.,
\begin{align}
q({\mathbf W}|{\mathbf Y};\widehat{\mathbf s}) =& \frac{1}{(\pi^{||\widehat s||_0}\det(\widehat{\mathbf C}_{\widehat{\mathcal S},0}))^{L}}\exp\left[-{\rm tr}\left(({\mathbf W}_{\widehat{\mathcal S}} - \widehat{\mathbf W}_{\widehat{\mathcal S}})^{\rm H}\widehat{\mathbf C}_{\widehat{\mathcal S},0}^{-1}({\mathbf W}_{\widehat{\mathcal S}} - \widehat{\mathbf W}_{\widehat{\mathcal S}})\right)\right]\prod_{i\not\in\widehat{\mathcal S}}\delta(\mathbf w_i)\label{qwpdf}\\
=&\prod\limits_{l=1}^L{\mathcal {CN}}({\mathbf w}_{\widehat{\mathcal S},l};\widehat{\mathbf w}_{\widehat{\mathcal S},l},\widehat{\mathbf C}_{\widehat{\mathcal S},0})\prod_{i\not\in\widehat{\mathcal S}}\delta(w_{i,l}),\label{qwpdfiid}
\end{align}
where ${\mathbf w}_{\widehat{\mathcal S},l}$ denotes the $l$th column of $\widehat{\mathbf W}_{\widehat{\mathcal S}}$.
From (\ref{qwpdfiid}), it can be seen that each column of $\mathbf W_{\widehat{\mathcal S}}$ is independent and is a complex Gaussian distribution. This is convenient for parallel execution, as described in Section \ref{reat}.

To calculate $q({\mathbf W}|{\mathbf Y})$, $\widehat{\mathbf s}$ has to be given. Plugging the postulated PDF $q({\boldsymbol \Phi}|{\mathbf Y})$ (\ref{postpdf}) in (\ref{DL-L}), one has
\begin{align}\notag
&\ln Z(\widehat{\mathbf s})\triangleq {\mathcal L}(q(\boldsymbol\theta\mathbf {,W,s|Y});\widehat{\mathbf s})= {\rm E}_{q(\boldsymbol\theta,{\mathbf W},{\mathbf s}|{\mathbf Y})}\left[\ln\tfrac{p({\mathbf Y}, \boldsymbol\theta,{\mathbf W},{\mathbf s};\widehat{\mathbf s})}{q({\boldsymbol\theta\mathbf {,W,s|Y}};\widehat{\mathbf s})}\right]\notag\\
=& {\rm E}_{q(\boldsymbol\theta,{\mathbf W},{\mathbf s}|{\mathbf Y})}[\sum_{i=1}^N \ln p(s_i) + \ln p({\mathbf W}|{\mathbf s})+\ln p({\mathbf Y}|\boldsymbol \theta,\mathbf W)- \ln q({\mathbf W}|{\mathbf Y})]+{\rm const}\notag\\
=&-L\ln\det(\mathbf J_{\widehat{\mathcal S}}+\frac{\nu}{\tau}\mathbf I_{|\widehat{\mathcal S}|})+||\widehat{\mathbf s}||_0\ln\frac{\lambda}{1-\lambda}+\nu^{-1}{\rm tr}(\mathbf H_{\widehat{\mathcal S}}^{\rm H}(\mathbf J_{\widehat{\mathcal S}}+\frac{\nu}{\tau}{\mathbf I}_{|\widehat{\mathcal S}|})^{-1}{\mathbf H}_{\widehat{\mathcal S}})+||\widehat{\mathbf s}||_0L\ln\frac{\nu}{\tau}+{\rm const}.\label{lnZ}
\end{align}
Thus $\widehat{\mathbf s}$ should be chosen to maximize $\ln Z(\widehat{\mathbf s})$ (\ref{lnZ}).

The computation cost of enumerative method to find the globally optimal binary sequence $\mathbf s$ of (\ref{lnZ}) is $O(2^N)$, which is impractical for typical values of $N$. Here a greedy iterative search strategy similar to \cite{Badiu} is proposed. For a given $\widehat{\mathbf s}$, we update it as follows: For each $k=1,\cdots,N$, calculate $\Delta_k=\ln Z(\widehat{\mathbf s}^k)-\ln Z(\widehat{\mathbf s})$, where $\widehat{\mathbf s}^k$ is the same as $\widehat{\mathbf s}$ except that the $k$th element of $\widehat{\mathbf s}$ is flipped. Let $k^*=\underset{k}{\operatorname {argmax}}~\Delta_k$. If $\Delta_{k^*}>0$, we update $\widehat{\mathbf s}$ with the $k^*$th element flipped, and $\widehat{\mathbf s}$ is updated, otherwise $\widehat{\mathbf s}$ is kept, and the algorithm is terminated. In fact, $\Delta_k$ can be easily calculated and the details are provided in Appendix \ref{modelSelection}.

Since each step increases the objective function (which is bounded) and $\mathbf s$ can take a finite number of values (at most $2^N$), the method converges in a finite number of steps to some local optimum. If deactive is not allowed and $\hat{\mathbf s}^0$ is initialized as ${\mathbf 0}_N$, then it can be proved that finding a local maximum of $\ln Z(\widehat{\mathbf s})$ costs only $O(\hat{K})$ steps. In general, numerical experiments show that $O(\hat{K})$ steps is often enough to find the local optimum.
\subsection{Estimating the model parameters}
After updating the frequencies and weights, the model parameters $\boldsymbol\beta = \{\nu,~\lambda,~\tau\}$ is estimated via maximizing the lower bound ${\mathcal L}(q({\boldsymbol \Phi}|{\mathbf Y});{\boldsymbol \beta})$ for fixed $q({\boldsymbol\Phi}|{\mathbf Y})$. In Appendix \ref{estmopa}, it is shown that
\begin{align}
&{\mathcal L}(q{(\boldsymbol\theta\mathbf {,W,s|Y})};\boldsymbol \beta)= {\rm E}_{q{(\boldsymbol\theta\mathbf {,W,s|Y}})}\left[\ln{\tfrac{p({\mathbf {Y,}\boldsymbol\theta\mathbf{,W,s};\boldsymbol \beta})}{q({\boldsymbol\theta\mathbf {,W,s|Y}})}}\right]\notag\\
=&-\frac{1}{\nu}\left[||\mathbf Y-{\widehat{\mathbf A}}_{\widehat{\mathcal S}}{\widehat{\mathbf W}}_{\widehat{\mathcal S}}||^2_{\rm F}+L{\rm tr}(\mathbf J_{\widehat{\mathcal S}}\widehat{{\mathbf C}}_{\widehat{\mathcal S},0})\right]-\frac{1}{\tau}[{\rm tr}(\widehat{\mathbf W}_{\widehat{\mathcal S}}\widehat{\mathbf W}^{\rm H}_{\widehat{\mathcal S}})+L{\rm tr}(\widehat{{\mathbf C}}_{\widehat{\mathcal S},0})]\notag\\
+& ||\widehat{\mathbf s}||_0(\ln\frac{\lambda}{1-\lambda}-L{\rm ln}\tau)+N\ln(1-\lambda)-ML{\rm ln}\nu+{\rm const}.\label{Lvaluemodpa}
\end{align}
Setting $\frac{\partial\mathcal L}{\partial\nu}=0$, $\frac{\partial\mathcal L}{\partial\lambda}=0$, $\frac{\partial\mathcal L}{\partial\tau}=0$, we have
\begin{align}\label{mu-rou-tau-hat}
\widehat{\nu}& = {||\mathbf Y-{\widehat{\mathbf A}}_{\widehat{\mathcal S}}{\widehat{\mathbf W}}_{\widehat{\mathcal S}}||^2_{\rm F}}/({ML})+{{\rm tr}(\mathbf J_{\widehat{\mathcal S}}\widehat{{\mathbf C}}_{\widehat{\mathcal S},0})}/{M}+\sum_{i\in\widehat{\mathcal S}}\sum_{l=1}^L|\widehat{W}_{il}|^2(1-{||\widehat{\mathbf a}_i||_2^2}/M)/L,\notag\\
\widehat{\lambda}  &=\frac{||\widehat{\mathbf s}||_0}{N},\quad \quad \widehat{\tau} = \frac{{\rm tr}(\widehat{\mathbf W}_{\widehat{\mathcal S}}\widehat{\mathbf W}^{\rm H}_{\widehat{\mathcal S}})+L{\rm tr}(\widehat{{\mathbf C}}_{\widehat{\mathcal S},0})}{L||\widehat{\mathbf s}||_0}.
\end{align}
\subsection{The MVALSE algorithm}
Now the details of updating the assumed posterior $q({\boldsymbol \theta},{\mathbf W},{\mathbf s}|{\mathbf Y})$ have been given and summarized in Algorithm \ref{VALSEMMV}. For the proposed algorithm, the initialization is important for the performance of the algorithm. The schemes that we initialize $\widehat{\nu}$, $\widehat{\lambda}$, $\widehat{\tau}$ and $q{(\theta_i|\mathbf Y)}$, $i\in\{1,\cdots,N\}$ are below.

First, initialize $q({\theta_1|{\mathbf Y}})$ as $q({\theta_1|{\mathbf Y}})\propto {\rm exp}\left(\frac{||{\mathbf Y}^{\rm H}{\mathbf a}(\theta_1)||^2_2}{\nu M}\right)$, which can be simplified as the form similar to (\ref{pdf-q}): By defining ${\mathcal M}' = \{m-n~|~m,n\in\{0,1,\cdots,M-1\},m>n\}$ with cardinality $M'=M-1$ and ${\mathbf a}':[-\pi,\pi)\to{\mathbb C}^{M'},\theta \to {\mathbf a}'(\theta) \triangleq ({\rm e}^{{\rm j}\theta m}~|~m\in{\mathcal M}')^{\rm T}$. Obviously ${\mathbf a}(\theta)=[1;{\mathbf a}'(\theta)]$. For each $t = 1,\cdots,M'$, by constructing $\gamma_t$ as $ \gamma_t= \frac{1}{M}\sum_{(k,l)\in{\mathcal T}_t}{\mathbf Y}_{k,:}{\mathbf Y}^{\rm H}_{l,:}$ with ${\mathcal T}_t = \{(k,l)~|~1\leq k,l\leq M,m_k-m_l=t\}$, $q({\theta_i|\mathbf Y})$ can be re-expressed as
\begin{align}\label{INI-MVM}
q({\theta_i}|{\mathbf Y})\propto{\rm exp}\left({\rm Re}\left\{\frac{2}{\nu}\gamma^{\rm H}{\mathbf a}'(\theta_1)\right\}\right).
\end{align}
Then $\hat{\mathbf a}_1={\rm E}[{\mathbf a}(\theta_1)]$ can be calculated. Since $J_1=M$ and ${\mathbf H}_1$ (\ref{J-H}) can be calculated. According to (\ref{defCshat0}) and (\ref{What}), $\hat{\mathbf w}_1$ is calculated. Then we update $q({\theta_2}|{\mathbf Y})\propto {\rm exp}\left(\frac{||{\mathbf Y}_1^{\rm H}{\mathbf a}(\theta_1)||^2_2}{\nu M}\right)$ with ${\mathbf Y}_1={\mathbf Y} - \hat{\mathbf a}_1\hat{\mathbf w}_1^{\rm T}$. Following the previous steps, $q(\theta_i|{\mathbf Y})$, $\hat{\mathbf a}_i$ and $\hat{\mathbf w}_i$ are all initialized. As for the model parameters $\boldsymbol \beta$, ${\boldsymbol \gamma}=[\gamma_1,\cdots,\gamma_{M-1}]^{\rm T}\in {\mathbb C}^{M-1}$ is used to build a Toeplitz estimate of ${\rm E}[{\mathbf Y}{\mathbf Y}^{\rm H}]$. Let $L\hat\nu$ be the average of the lower quarter of the eigenvalues of that matrix, and $\hat\tau$ is initialized as $\hat\tau = ({\rm tr}[\mathbf Y^{\rm H}\mathbf Y]/M-L\hat{\nu})/(\hat{\rho}N)$, where the active probability $\rho$ is initialized as $\hat{\rho} = 0.5$.

The complexity of MVALSE algorithm is dominated by the two steps \cite{Badiu}: the maximization of $\ln Z({\mathbf s})$ and the approximations of the posterior PDF $q(\theta|{\mathbf Y})$ by mixtures of von Mises PDFs. For the maximization of $\ln Z({\mathbf s})$, if $\mathcal S$ is initialized such that $|\hat{\mathcal S}| = 0$ and deactive is not allowed, it can be proved that the greedy iterative search strategy needs at most $N$ steps to converge. For the general case where deactive is allowed, numerical experiments show that $O(N)$ steps is enough to converge. For each step, the computational complexity is $O(N^2+NL)$ due to the matrix multiplication. Therefore, the computational complexity is $O(N^4+N^3L)$. For the approximations of the posterior PDF $q(\theta|{\mathbf Y})$ by mixtures of von Mises PDFs, the Heuristic $2$ method \cite[subsection D of Section IV]{Badiu} is adopted and the computational complexity is $O(N^2M + M^2N + N^2L+MNL)$. In conclusion, the dominant computational complexity of the MVALSE is $O[(N^4+N^3L)\times T]$ with $T$ being the number of iterations as $M$ is close to $N$.
\begin{algorithm}[ht]
\caption{Outline of MVALSE algorithm with MMVs setting.}\label{VALSEMMV}
\textbf{Input:}~~Signal matrix $\mathbf Y$\\
\textbf{Output:}~~The model order estimate $\widehat{K}$, frequencies estimate $\widehat{\boldsymbol\theta}_{\widehat{\mathcal S}}$, complex weights estimate $\widehat{\mathbf W}_{\widehat{\mathcal S}}$ and reconstructed signal $\widehat{\mathbf X}$
\begin{algorithmic}[1]
\STATE Initialize $\widehat{\nu},\widehat{\lambda},\widehat{\tau}$~and~$q_{\theta_i|\mathbf Y},i\in\{1,\cdots,N\}$; compute $\widehat{\mathbf a}_i$
\STATE \textbf{repeat}
\STATE ~~~~Update~$\widehat{\mathbf s},\widehat{\mathbf W}_{\widehat{\mathcal S}}~{\rm and}~\widehat{\mathbf C}_{\widehat{\mathcal S}}$ ({\rm Sec}.\ref{W-and-C})
\STATE ~~~~Update~$\widehat{\nu}$, $\widehat{\lambda}$, $\widehat{\tau}$ (\ref{mu-rou-tau-hat})
\STATE ~~~~Update~$\boldsymbol\eta_i$~and~$\widehat{\mathbf a}_i$ for all $i\in \widehat{\mathcal S}$ ({\rm Sec}.\ref{yita})
\STATE \textbf{until} stopping criterion is satisfied
\STATE \textbf{return} ${\widehat K}$, $\widehat{\boldsymbol\theta}_{\widehat{\mathcal S}}$, $\widehat{\mathbf W}_{\widehat{\mathcal S}}$ and $\widehat{\mathbf X}$
\end{algorithmic}
\end{algorithm}
\section{MVALSE with Parallel Processing}\label{reat}
The MVALSE Algorithm \ref{VALSEMMV} is compared with the VALSE algorithm \cite{Badiu}. The MMVs can be decoupled as $L$ SMVs. For each SMV, we perform the VALSE algorithm and obtain ${\boldsymbol \eta}_{i,l}$ according to \cite[eq. (17)]{Badiu} for the $l$th snapshot, i.e.,
\begin{align}
{\boldsymbol \eta}_{i,l}=\frac{2}{\nu}\left({\mathbf y}_l-\sum_{j\in\widehat{\mathcal S} \backslash \{i\}}\widehat{\mathbf a}_j[{\widehat{\mathbf w}}_{j}^{\rm T}]_l\right)[{\widehat{\mathbf w}}_{i}^*]_l-\frac{2}{\nu}\sum_{j\in\widehat{\mathcal S} \backslash \{i\}}[{\widehat{\mathbf C}}_{j,i}]_{l,l}\widehat{\mathbf a}_j,
\end{align}
where $[{\widehat{\mathbf C}}_{j,i}]_{l,l}$ denotes the $(l,l)$th element of ${\widehat{\mathbf C}}_{j,i}$, $[{\widehat{\mathbf w}}_{j}^{\rm T}]_l$ denotes the $l$th element of ${\widehat{\mathbf w}}_{j}^{\rm T}$. From (\ref{yita-i}), ${\boldsymbol \eta}_{i}$ is the sum of ${\boldsymbol \eta}_{i,l}$ for all the snapshots, i.e., ${\boldsymbol \eta}_{i}=\sum\limits_{l=1}^L{\boldsymbol \eta}_{i,l}$, and now each ${\boldsymbol \eta}_{i,l}$ is updated as ${\boldsymbol \eta}_{i}$. We use ${\boldsymbol \eta}_{i}$ to obtain estimates $\widehat{\theta}_i$ and $\widehat{\mathbf a}_i$ \cite{Badiu}. In addition, we update the weights and their covariance (\ref{W-C-1}) by applying the SMV VALSE. Let $\widehat{\mathbf w}^{\rm T}_{\widehat{\mathcal S},l}$ be the estimated weights of the $l$th snapshot, the whole weight matrix $\widehat{\mathbf W}_{\widehat{\mathcal S}}$ (\ref{W-C-1}) can be constructed as $[\widehat{\mathbf w}^{\rm T}_{\widehat{\mathcal S},1};\cdots;\widehat{\mathbf w}^{\rm T}_{\widehat{\mathcal S},L}]$. It is worth noting that equation (\ref{qwpdfiid}) reveals that for different snapshots, the weight vectors are uncorrelated. Besides, the covariance of the weights for each snapshot is the same, which means that the common covariance of the weight can be fed to the SMV VALSE. For updating ${\mathcal S}$ under the active case, according to \cite[equ. (40)]{Badiu}, the changes $\Delta_{k,l}$ for the $l$th snapshot is
\begin{align}
\Delta_{k,l} = \ln\frac{v_k}{\tau} + \frac{|[{\mathbf u}_k]_l|^2}{v_k}+\ln\frac{\lambda}{1-\lambda}.
\end{align}
Thus (\ref{delta-k-active}) can also be expressed as
\begin{align}\label{delta-k-activel}
\Delta_{k} = \sum\limits_{l=1}^L\Delta_{k,l}-(L-1)\ln\frac{\lambda}{1-\lambda},
\end{align}
which can be viewed as a sum of the results $\Delta_{k,l}$ from the VALSE in SMVs, minus an additional constant term $(L-1)\ln\frac{\lambda}{1-\lambda}$. Similarly, for the deactive case, (\ref{delta-k-deactive}) can be viewed as a sum of the results (equation (44) in \cite{Badiu}) from the VALSE in SMVs, plus an additional constant term $(L-1)\ln\frac{\lambda}{1-\lambda}$. The additional constant terms can not be neglected because we need to determine the sign of (\ref{delta-k-active}) and (\ref{delta-k-deactive}) to update ${\mathcal S}$. For the $l$th snapshot, running the VALSE algorithm yields the model parameters estimates
\begin{align}
\widehat{\nu}_l& = {||{\mathbf y}_l-{\widehat{\mathbf A}}_{\widehat{\mathcal S}}[{\widehat{\mathbf W}}_{\widehat{\mathcal S}}]_{:,l}||^2}/M+{{\rm tr}(\mathbf J_{\widehat{\mathcal S}}\widehat{{\mathbf C}}_{\widehat{\mathcal S},0})}/{M}+\sum_{i\in\widehat{\mathcal S}}|\widehat{w}_{il}|^2(1-{||\widehat{\mathbf a}_i||_2^2}/M),\notag\\
\widehat{\tau}_l &= \frac{\|[\widehat{\mathbf W}_{\widehat{\mathcal S}}]_{:,l}\|^2+{\rm tr}(\widehat{{\mathbf C}}_{\widehat{\mathcal S},0})}{||\widehat{\mathbf s}||_0}.
\end{align}
According to (\ref{mu-rou-tau-hat}), model parameters estimates $\widehat{\nu}$ and $\widehat{\tau}$ are updated as the average of their respective estimates, i.e., $\widehat{\nu}=\sum\limits_{l=1}^L\widehat{\nu}_l/L$ and $\widehat{\tau}=\sum\limits_{l=1}^L\widehat{\tau}_l/L$, where $\widehat{\nu}_l$ and $\widehat{\tau}_l$ denote the estimate of the $l$th SMV VALSE, and $\widehat{\lambda}$ can be naturally estimated.
\section{MVALSE for sequential estimation (Seq-MVALSE)}
The previous MVALSE algorithm is designed to process a batch of data. In fact, MVALSE is very suitable for sequential estimation. We develop the Seq-MVALSE algorithm for sequential estimation, which is very natural as MVALSE outputs conjugate priors of the frequency. Suppose that the whole data ${\mathbf Y}=[{\mathbf Y}_{g_1},{\mathbf Y}_{g_2},\cdots,{\mathbf Y}_{g_G}]$ is partitioned into $G$ groups, where $g_1+g_2+\cdots+g_G=L$. For the first group with data ${\mathbf Y}_{g_1}$, we perform the MVALSE and obtain the posterior PDF of the frequencies. Then the posterior PDF of the frequencies can be viewed as the prior of the frequencies, and the MVALSE is performed with data ${\mathbf Y}_{g_2}$. Following the previous steps, Seq-MVALSE can be obtained for sequential estimation. The Seq-MVALSE is summarized as Algorithm \ref{SeqVALSEMMV}.
\begin{algorithm}[h]
\caption{Outline of Seq-MVALSE.}\label{SeqVALSEMMV}
\textbf{Input:}~~Signal matrix ${\mathbf Y}=[{\mathbf Y}_{g_1},{\mathbf Y}_{g_2},\cdots,{\mathbf Y}_{g_G}]$\\
\textbf{Output:}~~The model order estimate $\widehat{K}$, frequencies estimate $\widehat{\boldsymbol\theta}_{\widehat{\mathcal S}}$, complex weights estimate $\widehat{\mathbf W}_{\widehat{\mathcal S}}$ and reconstructed signal $\widehat{\mathbf X}$
\begin{algorithmic}[1]
\STATE Initialize $\widehat{\nu},\widehat{\lambda},\widehat{\tau}$~and~$q_{\theta_i|{\mathbf Y}_{g_1}},i\in\{1,\cdots,N\}$; compute $\widehat{\mathbf a}_i$ \
\FOR {$j=1,\cdots,G$ }
\STATE Run the MVALSE algorithm with data ${\mathbf Y}_{g_j}$, and output the posterior PDF $p({\boldsymbol \theta}|{\mathbf Y}_{g_j})$.
\STATE Set $p({\boldsymbol \theta}|{\mathbf Y}_{g_j})$ as the prior distribution of the next data group.
\ENDFOR
\STATE Return ${\widehat K}$, $\widehat{\boldsymbol\theta}_{\widehat{\mathcal S}}$, $\widehat{\mathbf W}_{\widehat{\mathcal S}}$ and $\widehat{\mathbf X}$
\end{algorithmic}
\end{algorithm}
\section{Numerical Simulation}\label{NS}
In this section, substantial numerical simulations are performed to substantiate the MVALSE algorithm. We define signal-to-noise ratio (SNR) as ${\rm SNR} \triangleq 10{\rm log}(||\mathbf A(\widetilde{\boldsymbol\theta})\widetilde{\mathbf W}^{\rm T}||_{\rm F}^2/||\widetilde{\mathbf U}||_{\rm F}^2)$ and the normalized mean square error (NMSE) of $\widehat{\mathbf X}$ and $\widehat{\boldsymbol\theta}$ are ${\rm NMSE}(\widehat{\mathbf X}) \triangleq 10{\rm log}(||\widehat{\mathbf X}-{\mathbf A}(\widetilde{\boldsymbol\theta})\widetilde{\mathbf W}^{\rm T}||_{\rm F}^2/||{\mathbf A}(\widetilde{\boldsymbol\theta})\widetilde{\mathbf W}^{\rm T}||_{\rm F}^2)$ and ${\rm NMSE}(\widehat{\boldsymbol\theta}) \triangleq 10{\rm log}(||\widehat{\boldsymbol\theta} - \widetilde{\boldsymbol\theta}||_2^2/||\widetilde{\boldsymbol\theta}||_2^2)$, the correct model order estimated probability $P(\widehat{K}=K)$ are adopted as the performance metrics. In the case when the model order is overestimated such that $\widehat{K}>K$, the top $K$ elements of $\widehat{\boldsymbol{\kappa}}$ is chosen to calculate the NMSE of the frequency, where $\widehat{\kappa}_i$ is the concentration parameter of the von Mises distribution approximated from the posterior $q(\theta_i|{\mathbf Y})$ (\ref{pdf-q}). When $\widehat{K}<K$, the frequencies are filled with zeros to calculated the NMSE of the frequency. The Algorithm \ref{VALSEMMV} stops when  $||\widehat{\mathbf X}^{(t-1)} - \widehat{\mathbf X}^{(t)}||_2/||\widehat{\mathbf X}^{(t-1)}||_2 < 10^{-5}$ or $t > 200$, where $t$ is the number of iteration.

In addition, the SPA method \cite{Yang1}, the Newtonized orthogonal matching pursuit (NOMP) method \cite{Mamandipoor, Lin} and the Cram\'er-Rao bound (CRB) derived in \cite{Lin} are chosen for performance comparison. For the SPA approach, the denoised covariance matrix is obtained firstly and the MUSIC method is used to avoid frequency splitting phenomenon, where the MUSIC method is provided by MATLAB \emph{rootmusic} and the optimal sliding window $W$ is empirically found. Here the sliding window $W$ is set as $W=12$. For the NOMP method, the termination condition is set such that the probability of model order overestimate is $1\%$ \cite{Lin}. All results are averaged over $10^3$ Monte Carlo (MC) trials unless stated otherwise.
\subsection{Performance investigation of MVALSE algorithm}
In this section, the performance of MVALSE algorithm is evaluated by varying SNR, the number of snapshots $L$ and the number of observations $M$. The frequencies are generated as follows: First, $K$ distributions are uniformly picked from $N$ von Mises distributions (\ref{prior_theta}) with $\mu_{0,i}=(2i-1-N)/(N+1)\pi$ and $\kappa_{0,i}=10^4$, $~i=1,\cdots,N$ without replacement. The frequencies $\{\theta_i\}_{i=1}^K$ are generated from the selected von Mises distribution and the minimum wrap-around distance is greater than $\Delta\theta=\frac{2\pi}{N}$. The elements of $\mathbf W$ are drawn i.i.d. from ${\mathcal {CN}}(1,0.1)$.  The wrap-around distance between any two generated frequencies is larger than $\Delta\theta=\frac{2\pi}{N}$. Other parameters are: $K=3$, $N=20$ \footnote{For the numerical experiment where $\kappa_{0,i}=10^4$ and $N=20$, straightforward calculation shows that the standard deviation of the von Mises distribution $\approx0.01$ and the distance between the adjacent frequencies is $\mu_{0,i+1}-\mu_{0,i}=0.3$. Thus the MVALSE with prior is almost a grid based method.}.
\subsubsection{Estimation by varying SNR}
\begin{figure*}
  \centering
  \subfigure[]{
    \label{SNR-X}
    \includegraphics[width=50mm]{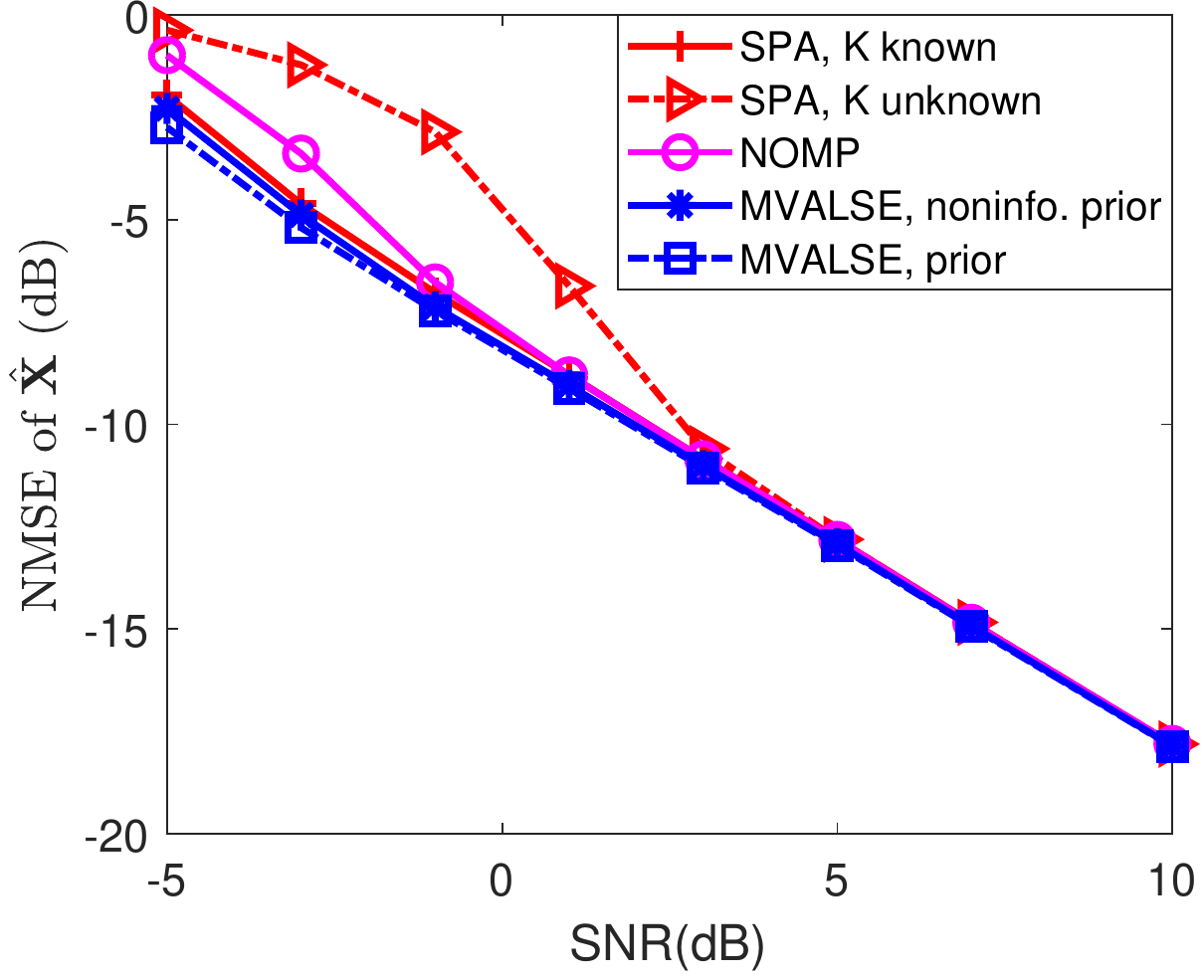}}
  \subfigure[]{
    \label{SNR-P}
    \includegraphics[width=50mm]{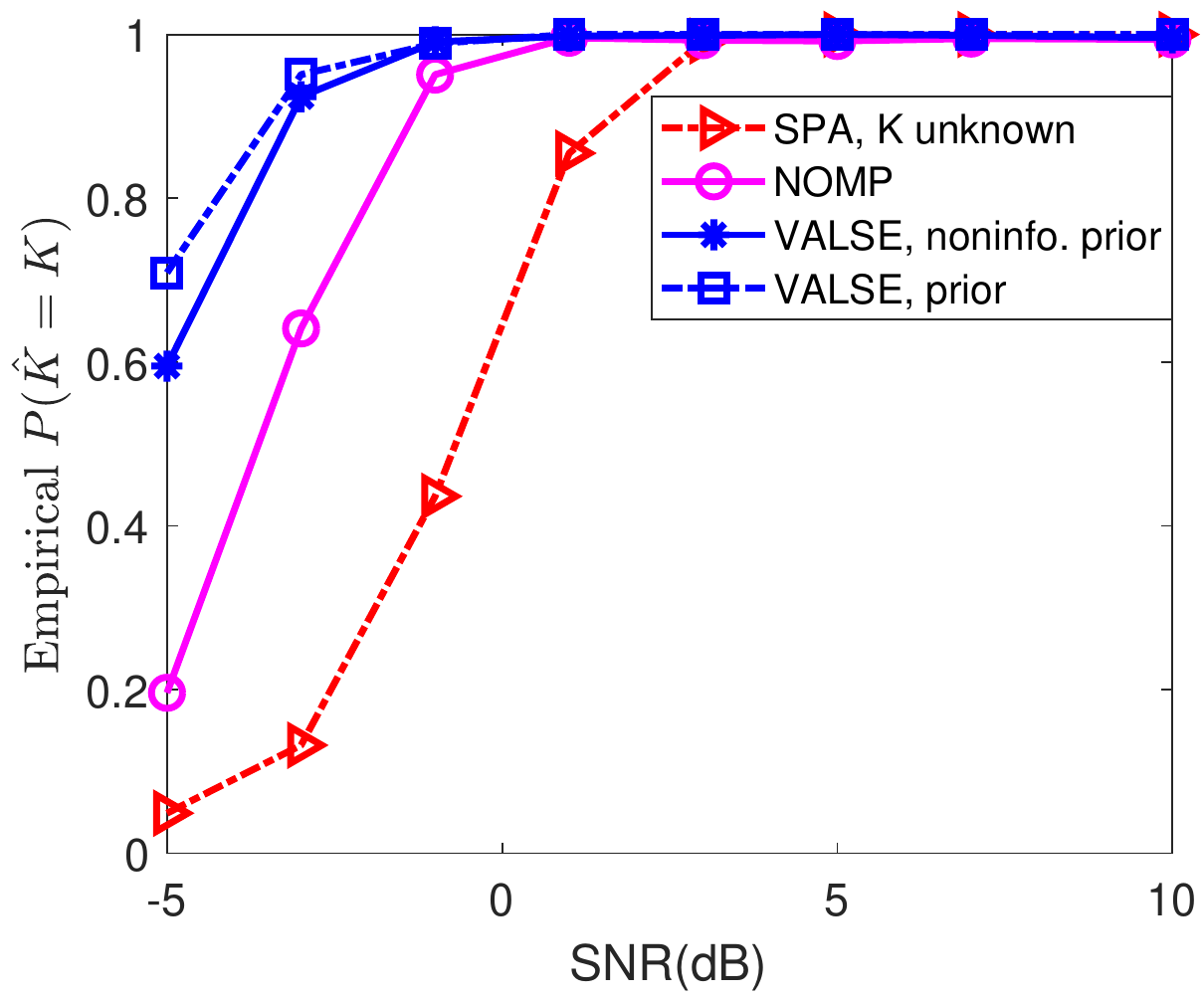}}
  \subfigure[]{
    \label{SNR-freq} 
    \includegraphics[width=50mm]{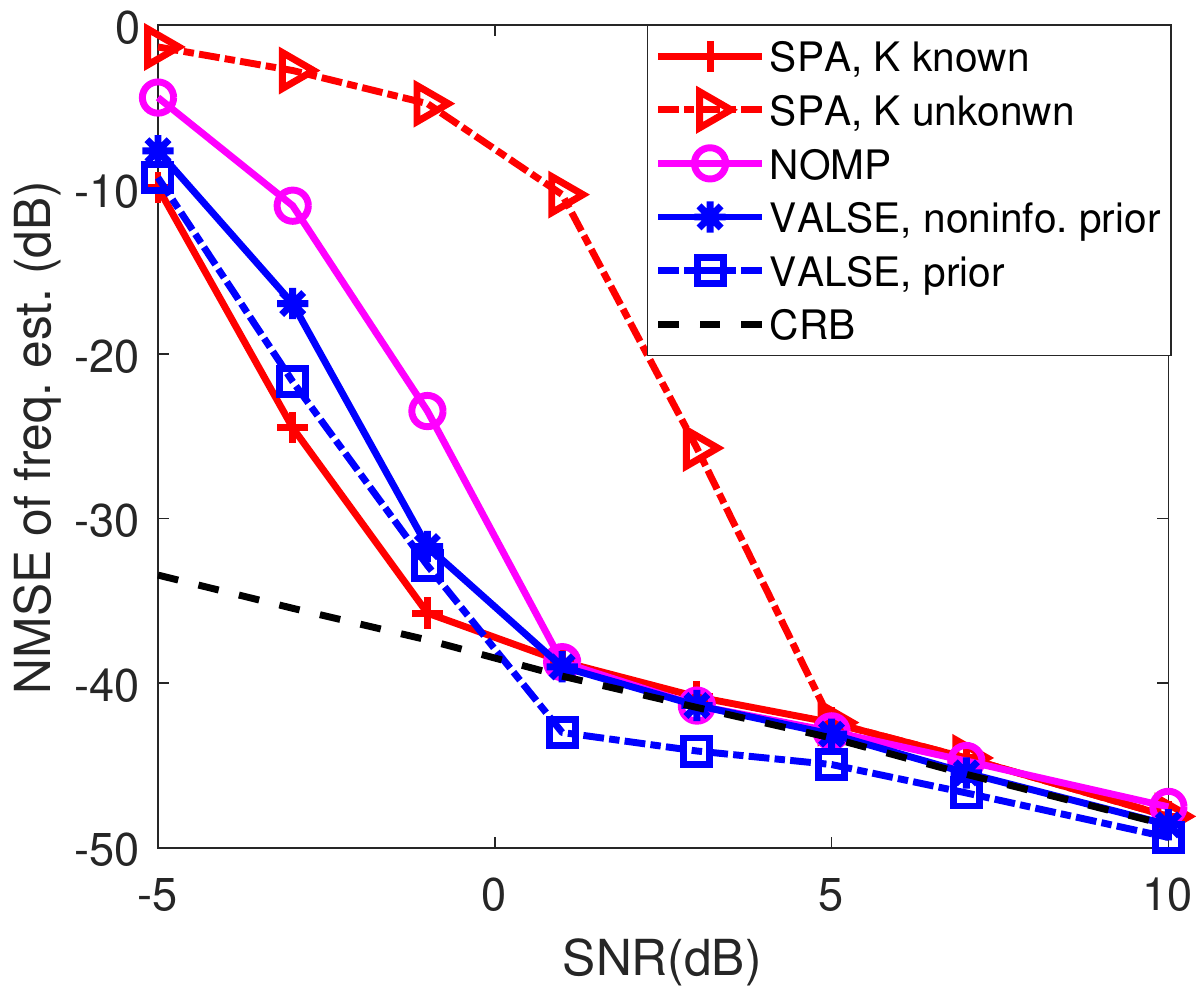}}
 \caption{Performance of algorithms by varying SNR. The number of measurements is $M=20$ and the number of snapshots is $L=8$. }
  \label{fig:subfig1} 
\end{figure*}

The performance in terms of model order estimation accuracy and frequency estimation error by varying SNR is presented in Fig. \ref{fig:subfig1}. We set the number of measurements $M = 20$ and snapshots $L = 8$. In Fig. \ref{SNR-X}, as the SNR increase, the NMSE of $\widehat{\mathbf X}$ decreases. When SNR $\geq3$ dB, the NMSEs of $\widehat{\mathbf X}$ are almost identical for all the algorithms. It can be seen that utilizing the prior information improves the performance of the VALSE algorithm. The frequency estimation error of the MVALSE with prior is smaller than the CRB , which makes sense because prior information is utilized. In Fig. \ref{SNR-P}, the VALSE algorithm achieves the highest probability of correct model order estimation, compared with NOMP and SPA algorithms.For the frequency estimation error, it is seen that the SPA (assuming $K$ is known ) approaches the CRB firstly. Then the VALSE and NOMP algorithms begin to approach the CRB. The SPA with unknown $K$ is the last one that approaches CRB.
\subsubsection{Estimation by varying L}
\begin{figure*}\label{L}
  \centering
  \subfigure[]{
    \label{L-X} 
    \includegraphics[width=50mm]{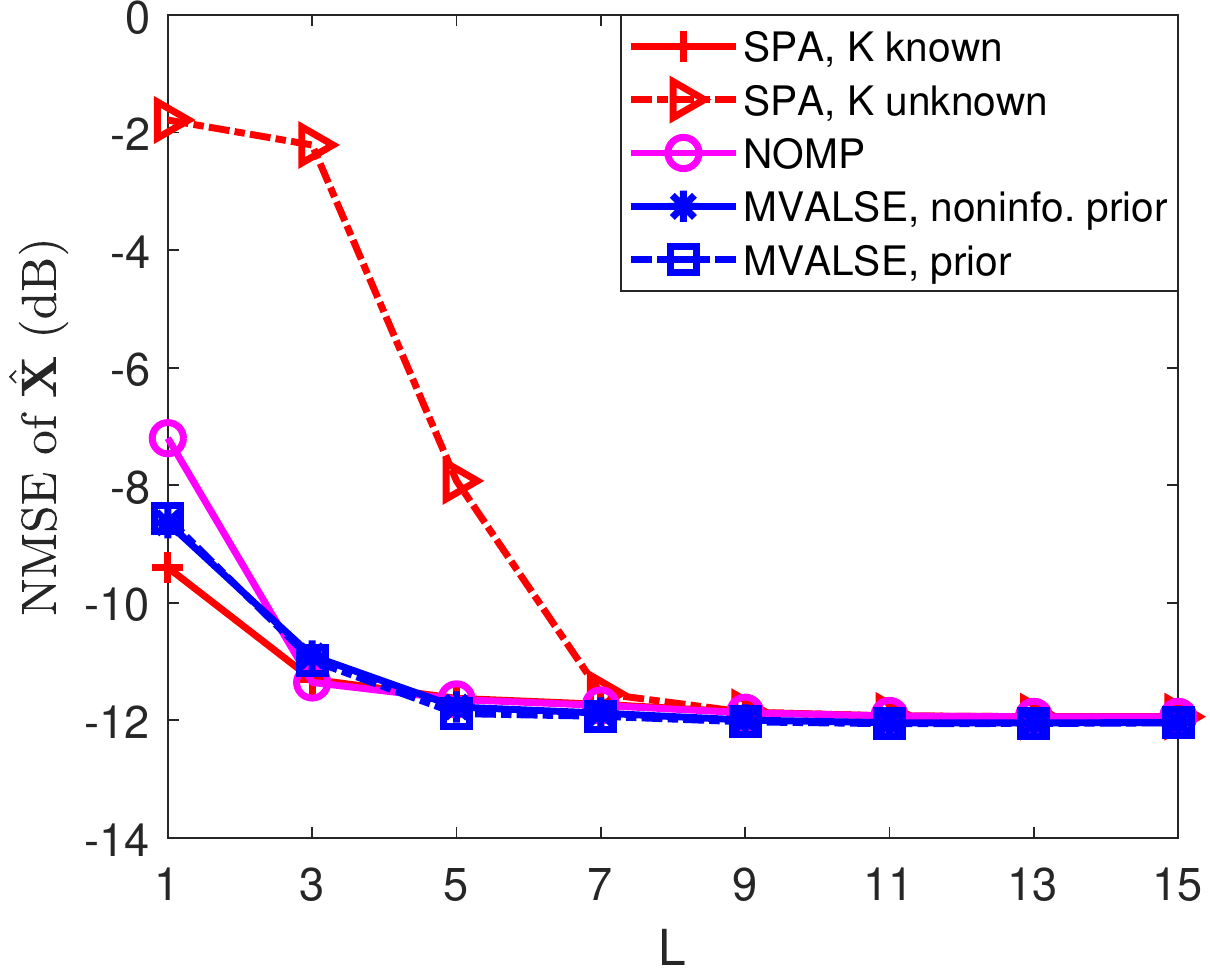}}
  \subfigure[]{
    \label{L-P} 
    \includegraphics[width=50mm]{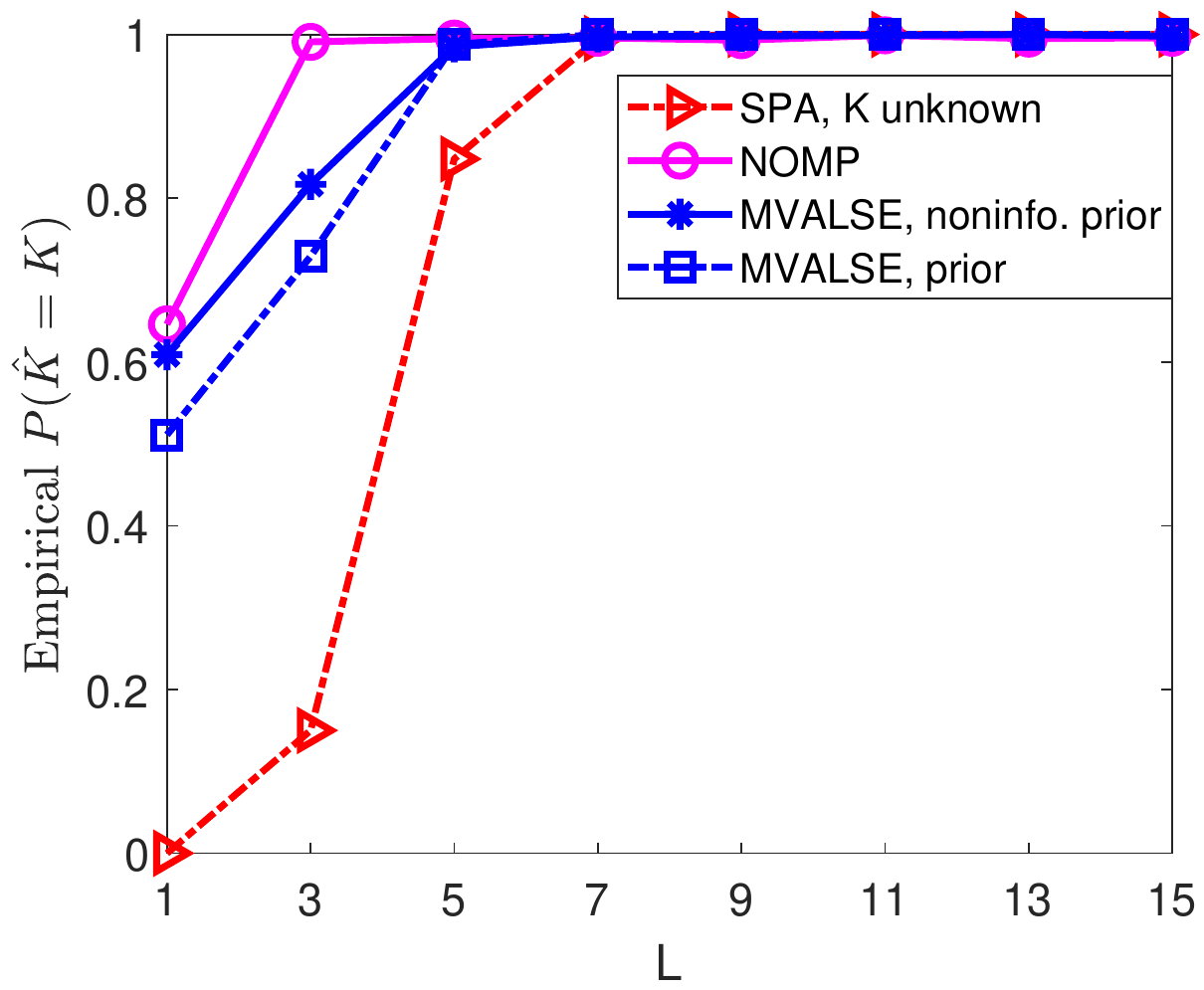}}
  \subfigure[]{
    \label{L-freq} 
    \includegraphics[width=50mm]{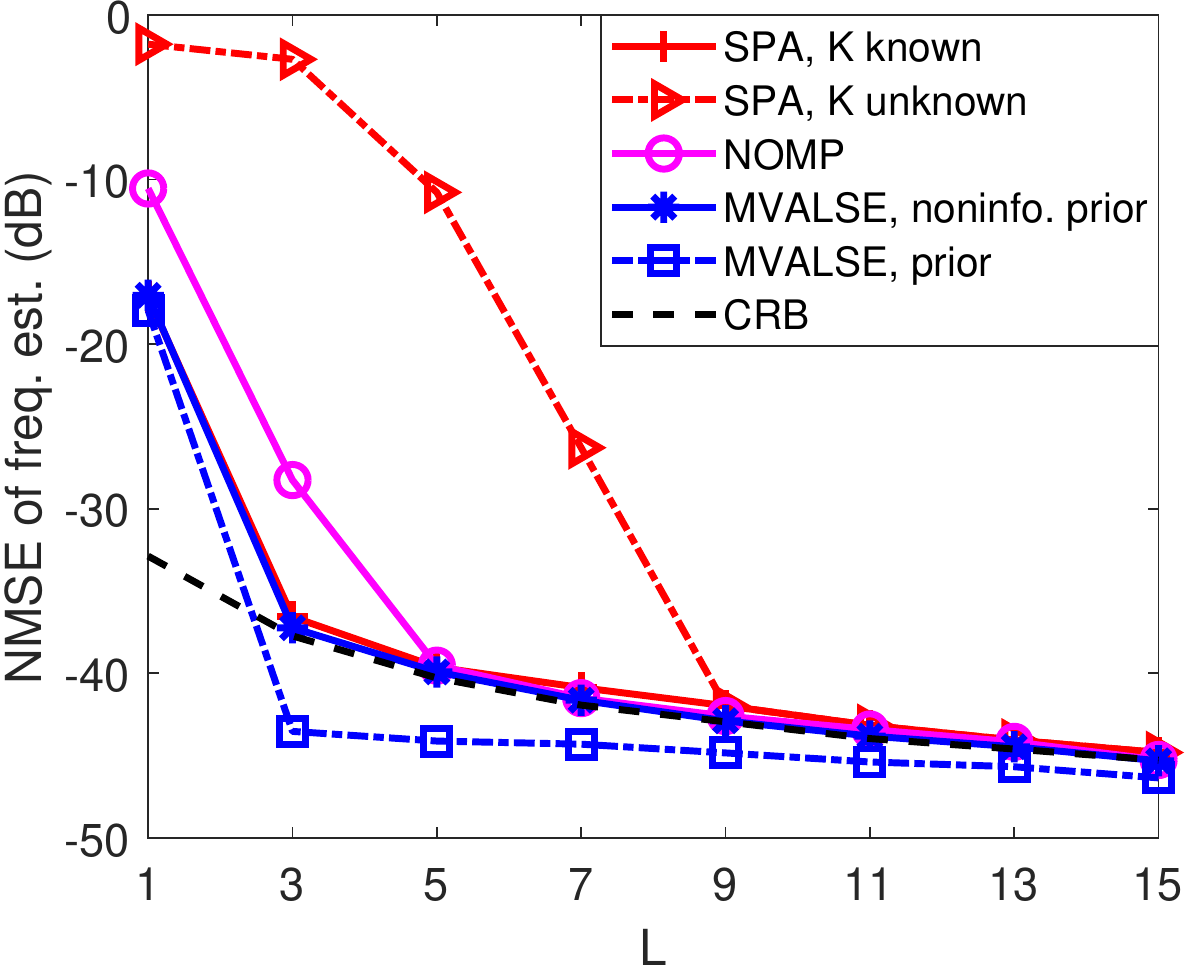}}
  \caption{Performance of algorithms by varying snapshots $L$. We set ${\rm SNR} = 0 {\rm dB}$ and the number of measurements $M=30$. }
  \label{fig:subfig2} 
\end{figure*}

In this subsection, we examine the estimation performance by varying the number of snapshots $L$. The number of measurements $M = 20$ and the $\rm{SNR}=4$ dB. The results are presented in Fig. \ref{fig:subfig2}. In Fig. \ref{L-X}, as the number of snapshots $L$ increases, the NMSE of $\widehat{\mathbf X}$ decreases and finally becomes stable. From Fig. \ref{L-P} and \ref{L-freq}, we can see that when $L\leq 3$, the NOMP algorithm achieves the highest probability of correct model order estimation, while its NMSE is higher than that of MVALSE methods. The reason is that the correct model order probability is not close to $1$, and the model order overestimate probability is only $1\%$, much smaller than the MVALSE methods shown in Table \ref{table2}. For the frequency estimation error in Fig. \ref{L-freq}, all the algorithms except the MVALSE with prior approach to the CRLB as $L$ increases. For the prior encoded MVALSE, its NMSE is lower than CRB.
\begin{table}[h!t]
    \begin{center}
\caption{The empirical probability of $\widehat{K}>K$  of the algorithms.}\label{table2}
    \begin{tabular}{|c|c|c|c|c|c|c|}
            \hline
            snapshots $L$&1&3&5&7\\ \hline
            MVALSE, prior & 33\% & 31\% & 1\%&0 \\ \hline
            MVALSE, noninfo. prior & 30\% & 23\% & 1\% & 0  \\ \hline
            NOMP & \multicolumn{4}{|c|}{1\%} \\ \hline
             \end{tabular}
    \end{center}
\end{table}
\subsubsection{Estimation by varying M}\label{M}
\begin{figure*}
  \centering
  \subfigure[]{
    \label{M-X} 
    \includegraphics[width=50mm]{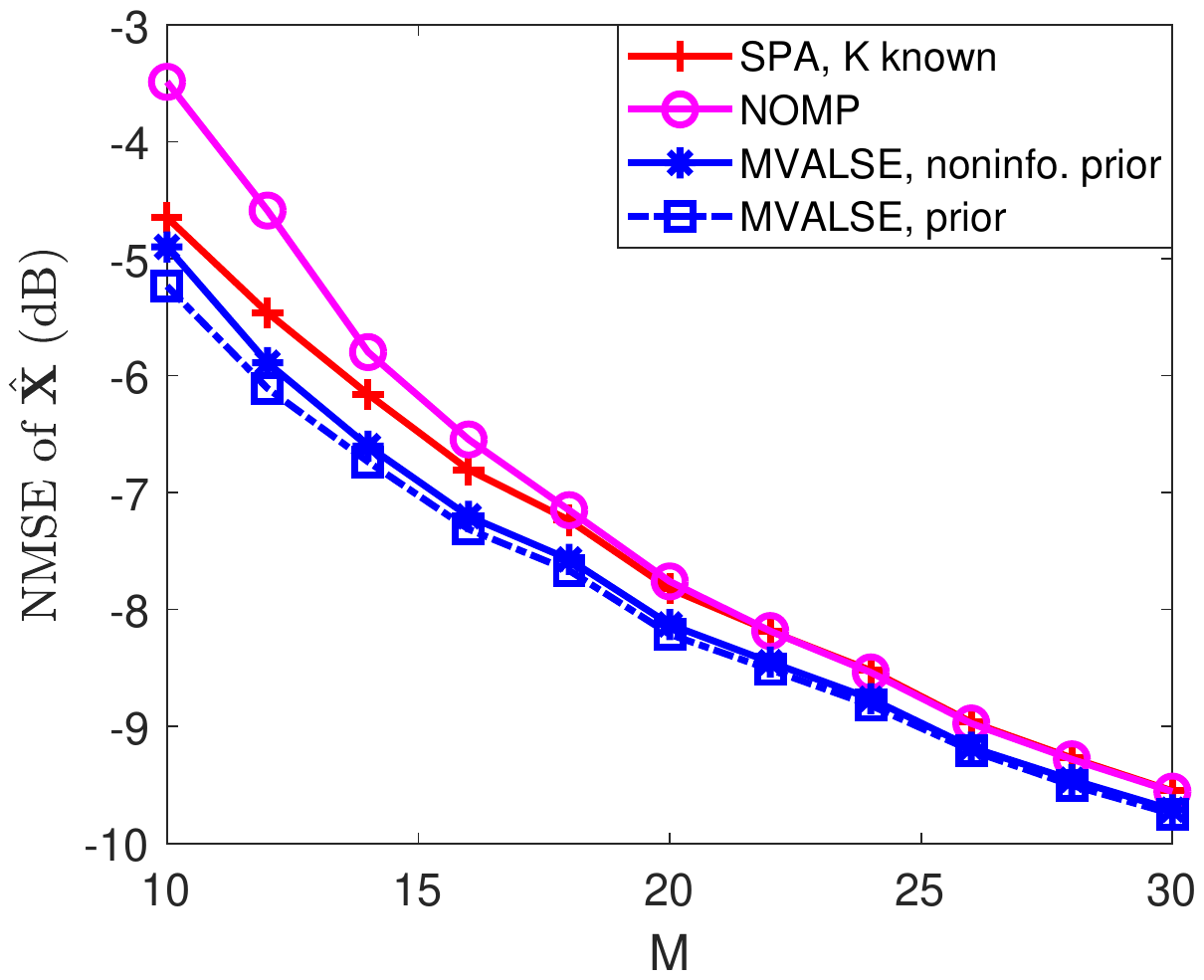}}
  \subfigure[]{
    \label{M-P} 
    \includegraphics[width=50mm]{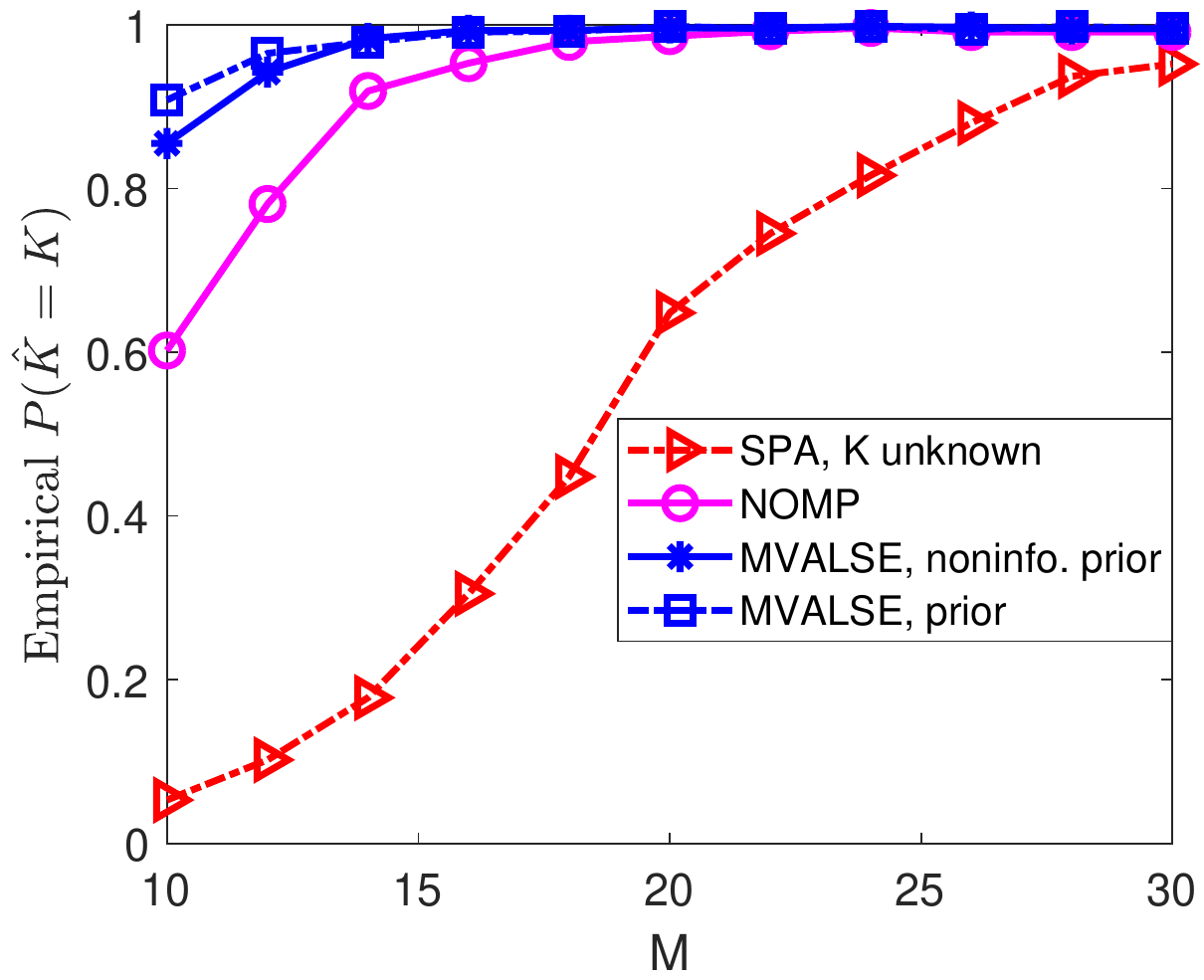}}
  \subfigure[]{
    \label{M-freq} 
    \includegraphics[width=50mm]{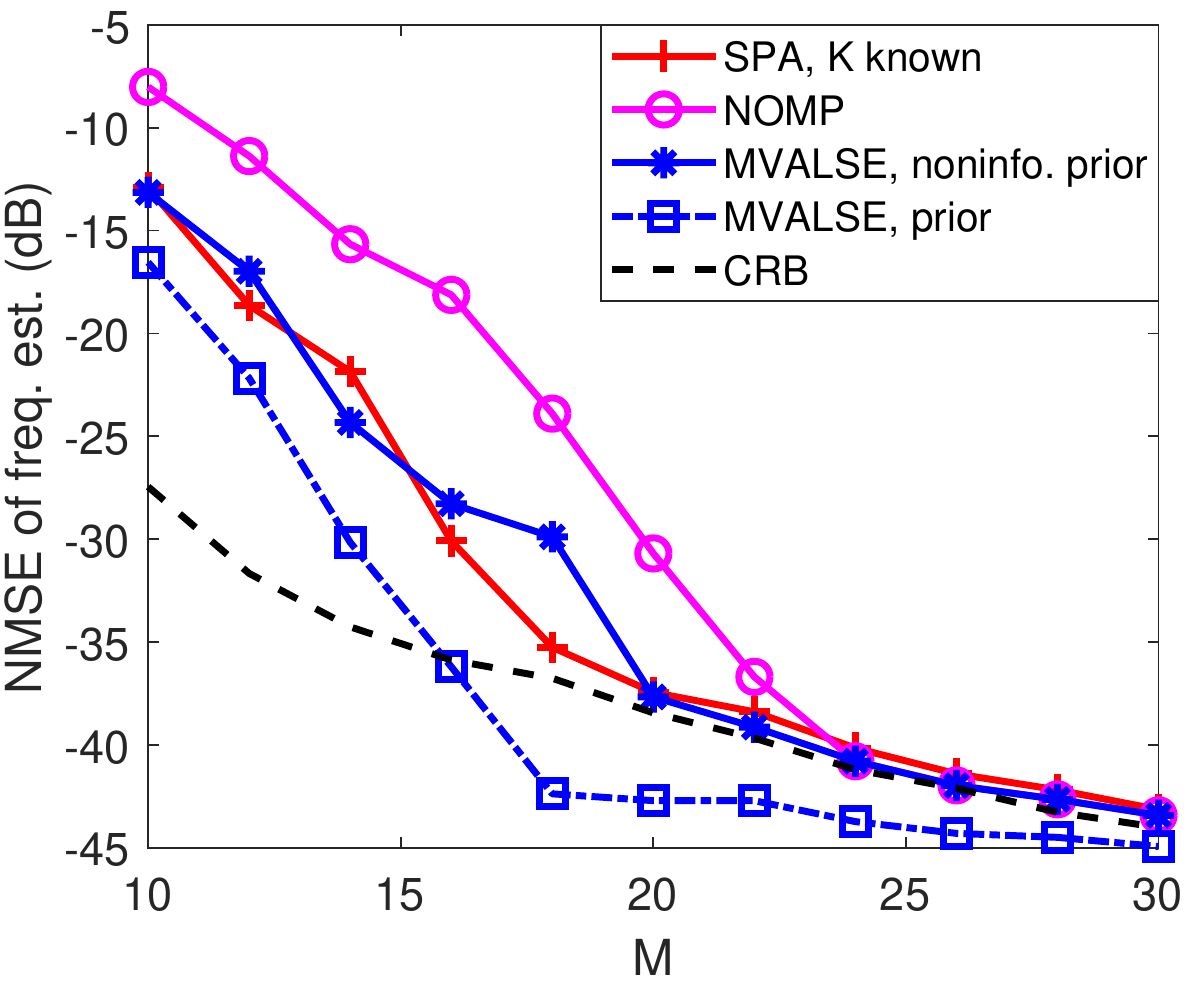}}
  \caption{The performance of algorithms by varying measurements $M$. We set ${\rm SNR}=0 $ dB and the number of snapshots $L = 8$. }
  \label{fig:subfig3} 
\end{figure*}

The performance is examined by varying the number of measurements per snapshots, and the results are presented in Fig. \ref{fig:subfig3}. For the first subfigure, the observations in Fig. \ref{SNR-X} and Fig. \ref{SNR-P} are also applicable in this scenario. The SPA with $K$ unknown in Fig. \ref{M-X} and Fig. \ref{M-freq} are not presented for the poor performance. In Fig. \ref{M-freq}, the MVALSE with prior performs best. SPA algorithm is the second algorithm that approach the CRB, and then MVALSE without prior follows. The NOMP algorithm approaches CRB lastly.
\subsection{Sequential estimation}
In this subsection, the performance of Seq-MVALSE is evaluated. The total number of snapshots is set as $L=8$. The snapshots are uniformly partitioned into $G$ groups. Here we investigate $G=1$, $G=4$ and $G=8$ groups, respectively. Note that performing MVALSE-S for $G=1$ is equivalent to performing the MVALSE. The frequencies are generated uniformly from $[-\pi, \pi]$. The wrap-around distance between any two frequencies is larger than $\Delta\omega=\frac{2\pi}{N}$. We set $K=3$, $M=20$ and $N=10$.

Two numerical experiments are conducted to investigate the performance of the Seq-MVALSE algorithm. For the first numerical experiment, the SNR is varied. It can be seen that as the SNR increases, the performances of all the algorithm improves. In addition, comparing the MVALSE algorithm, Seq-MVALSE has some performance degradation. As $G$ decreases, the performances of Seq-MVALSE improve. For the second numerical experiment, the performance is investigated with the whole number of snapshots fixed as $8$. It can be seen that the algorithm improves as the data arrives. For the fixed number of snapshots, the performance of Seq-MVALSE algorithm improves as $G$ decreases.
\begin{figure*}\label{SNR-S}
  \centering
  \subfigure[]{
    \label{SNR-S-X} 
    \includegraphics[width=50mm]{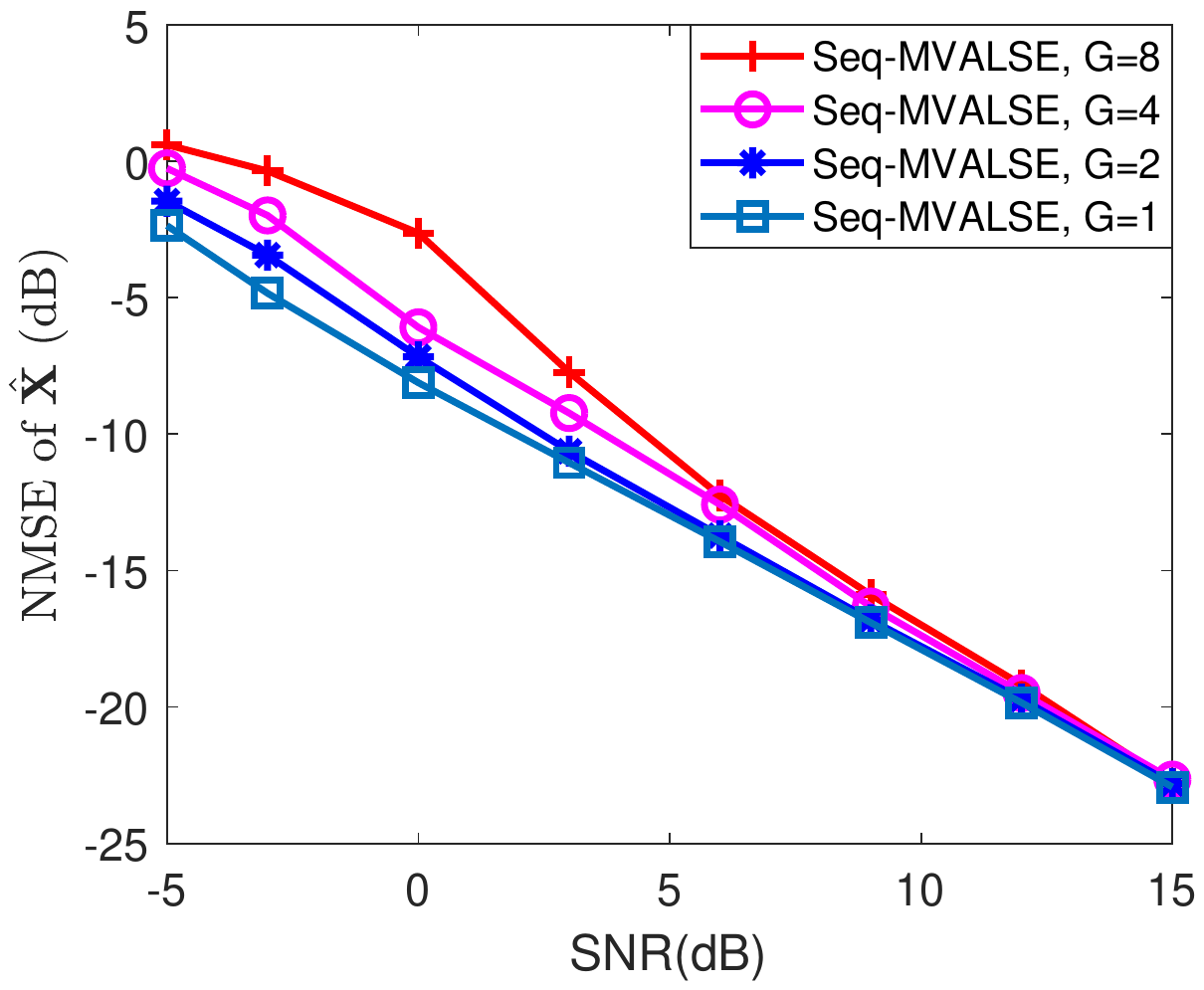}}
  \subfigure[]{
    \label{SNR-S-P} 
    \includegraphics[width=50mm]{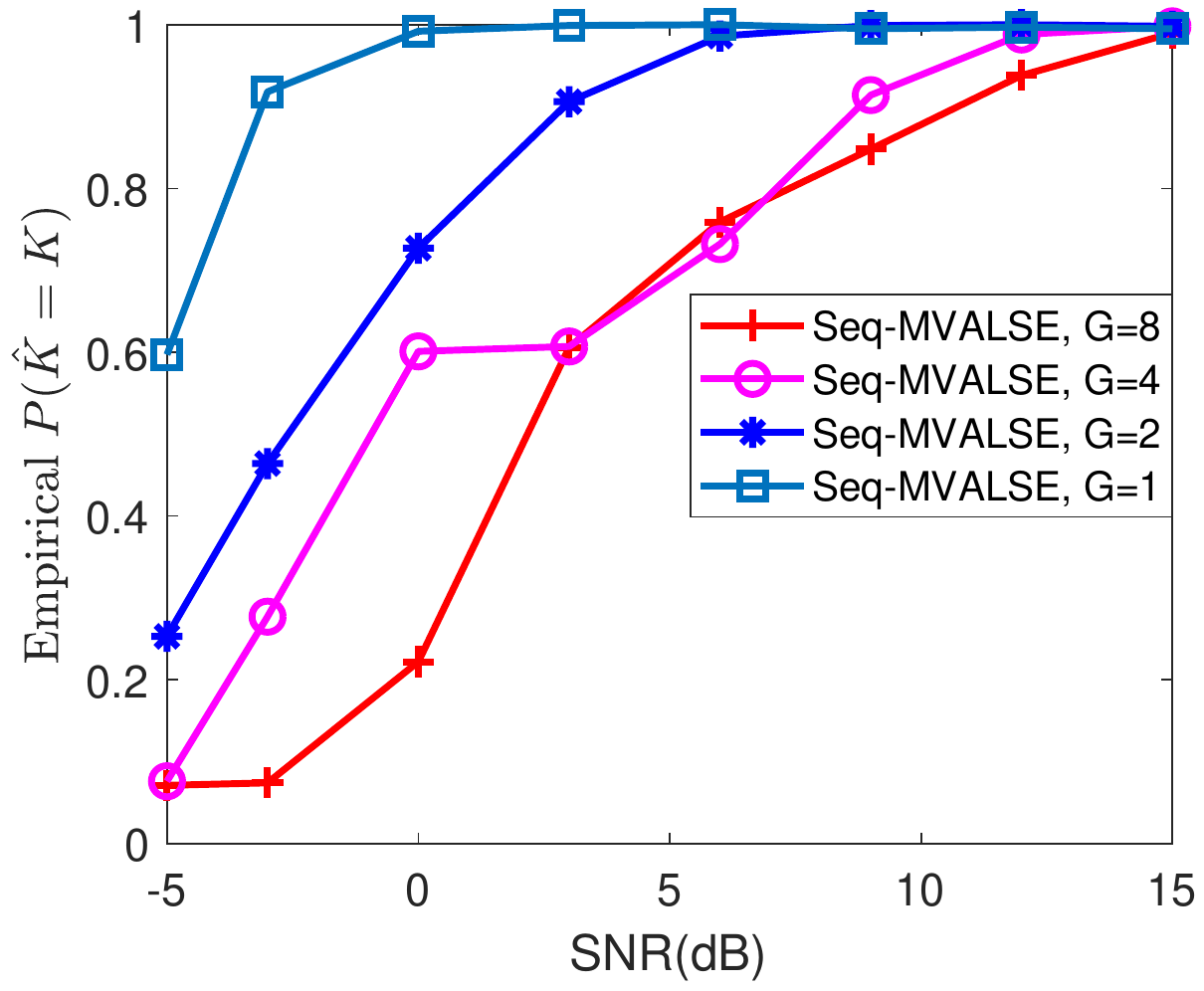}}
  \subfigure[]{
    \label{SNR-S-freq} 
    \includegraphics[width=50mm]{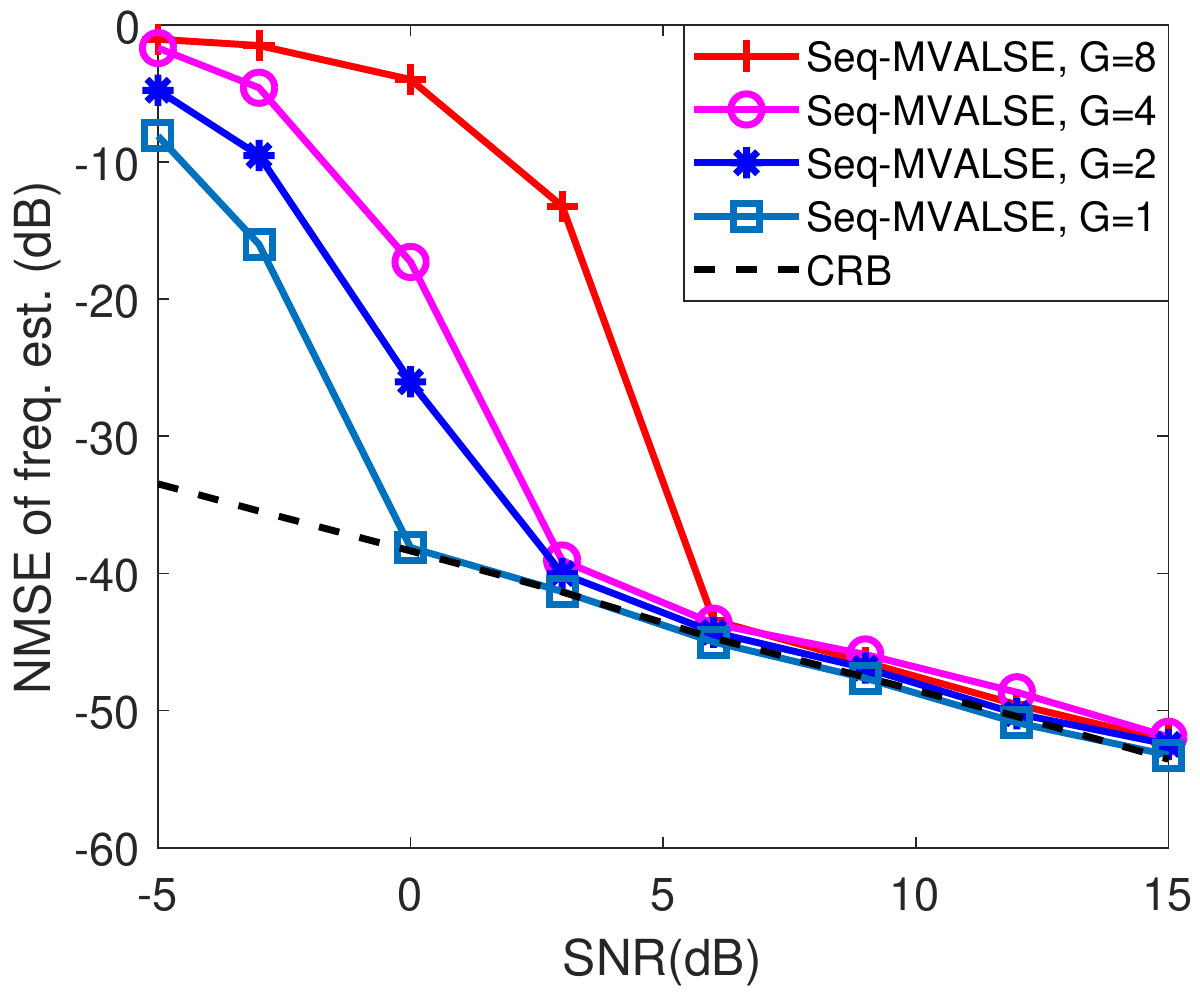}}
  \caption{Performance of MVALSE for sequential by varying SNR. We set $L = 8$ and the number of measurements $M=20$. }
\end{figure*}
\begin{figure}
  \centering
  \includegraphics[width=2.8in]{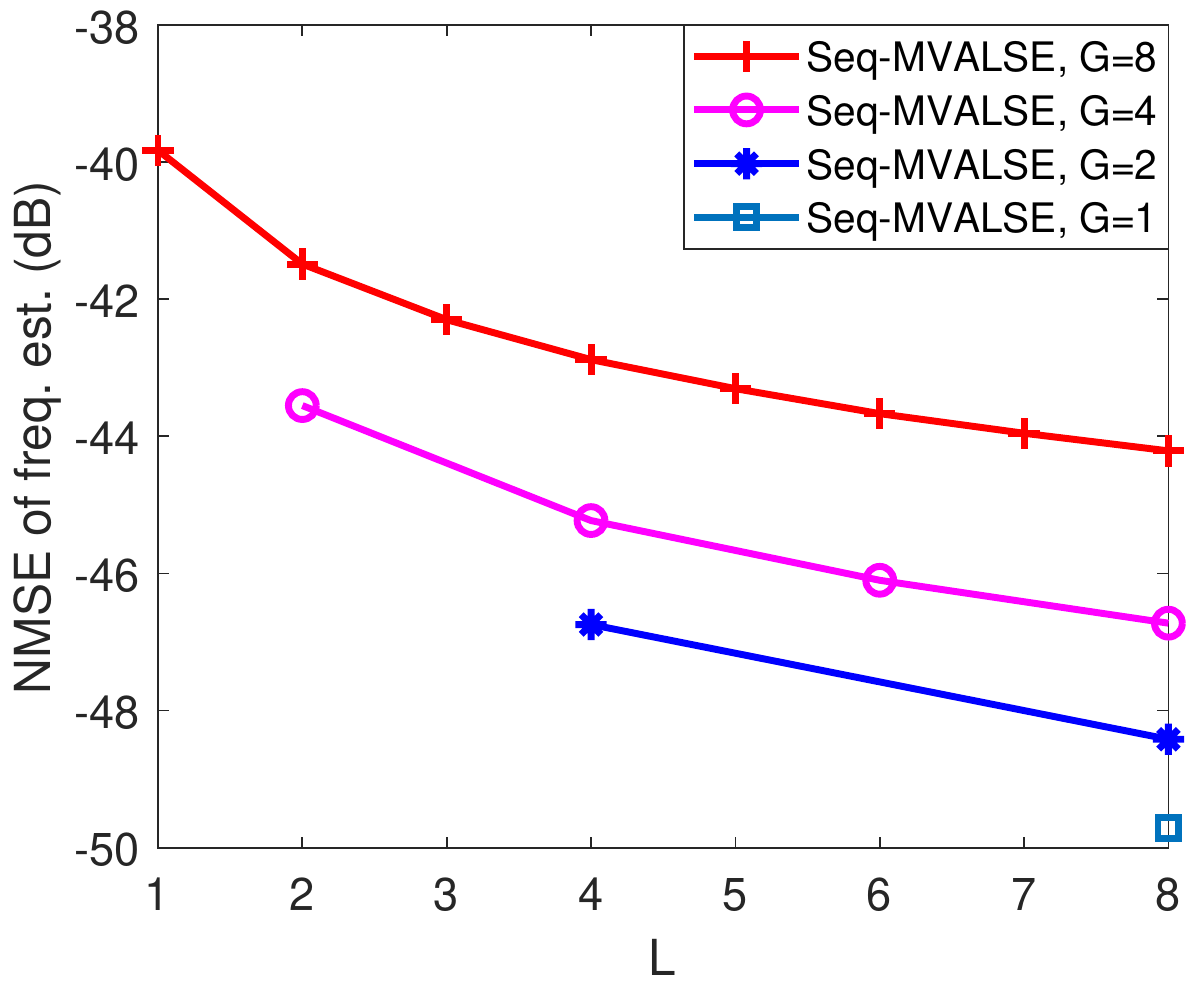}\\
  \caption{NMSE of frequency estimation of MVALSE for sequential by varying $L$. We set ${\rm SNR}=10$ dB and the number of measurements $M=20$. }  \label{L-S-freq}
\end{figure}
\subsection{Application: DOA Estimation}
\begin{figure}
  \centering
  \includegraphics[width=2.8in]{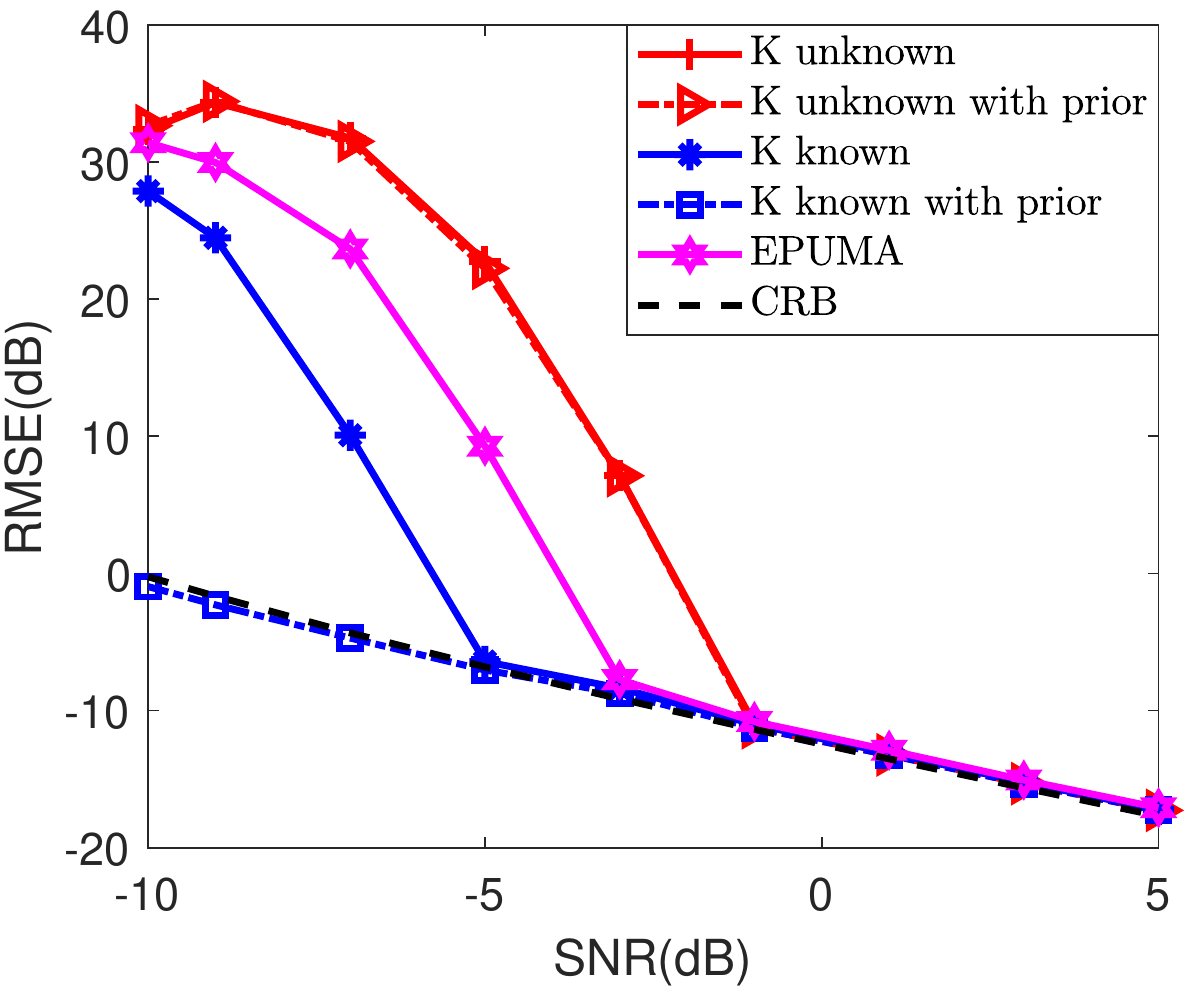}\\
  \caption{RMSE of MVALSE algorithm for DOA estimation. We set ${\rm SNR}=10$ dB and the number of measurements is $M=20$.}  \label{DOAres}
\end{figure}

The performance of MVALSE for DOA estimation is evaluated in this experiment. Let ${\boldsymbol \phi}\in {\mathbb R}^K$ denote the DOAs. For the DOA estimation problem where $K$ narrow band far-field signals impinging onto an $M$-element uniform linear array (ULA) whose interelement spacing $d$ is half of the wavelength $\lambda$, i.e., $d=\lambda/2$, the DOA estimation problem can be formulated as the LSE with ${\boldsymbol \theta}=\frac{2\pi d}{\lambda} {\rm sin}({\boldsymbol \phi})=\pi {\rm sin}({\boldsymbol \phi})$. We generate the frequencies $\boldsymbol \theta$ from the von Mises distribution, whose means corresponds to the DOAs $[5,9,70]^{\circ}$, and the concentration parameter is $\kappa_{0,i}=10^4$. We set $M=40$, $L=20$ and $K=3$. Since EPUMA approach outperforms many other subspace based DOA estimators, especially for small sample scenarios and provides reliable performance when the number of samples is small \cite{Huang}, we compare the MVALSE with EPUMA. Similar to \cite{Huang}, the root MSE (RMSE) ${\rm RMSE}\triangleq \sqrt{\sum\limits_{i=1}^K(\hat{\phi}_i-\phi_i)}$ is used to characterize the performance of the algorithms, where $\hat{\boldsymbol \phi}$ denotes the output of the algorithm. The results are presented in Fig. \ref{DOAres}. It can be seen that when $K$ is known, the MVALSE with prior always performs well. For the uninformative prior, the VALSE with known $K$ performs better than that of EPUMA \footnote{Given that $K$ is known, the VALSE is implemented without maximizing $\ln Z({\mathbf s})$.}. For $K$ unknown, the MVALSE with either prior or uninformative prior is worse than EPUMA. All these algorithm approach the CRB as SNR increases.

\section{Conclusion}\label{conclusion}
In this paper, the MVALSE algorithm is developed to jointly estimate the frequencies and weight coefficients in the MMVs setting. In contrast to related works which focuses on point estimates of the frequency, the MVALSE estimates the posterior PDF of the frequencies. It is also shown that the derived MVALSE is closely related to the VALSE algorithm, which is suitable for parallel processing. In addition, the performance of the MVALSE method with von Mises prior PDFs for the frequencies is studied. Furthermore, the MVALSE is extended to perform sequential estimation. Finally, substantial experiments are conducted to illustrate the competitive performance of the MVALSE method and its application to DOA problems, compared to other approaches. As for future work, referring to the unified inference framework proposed in \cite{Meng}, the MVALSE algorithm can be extended to solve the nonlinear measurement model, such as quantization \cite{offVALSEq}, off-grid millimeter wave channel estimation, phase retrieval and so on.
\section{Appendix}
\subsection{Derivation of ${q}({\theta}_i|{\mathbf Y})$}\label{derqtheta}
Substituting (\ref{ahat}) and (\ref{w_est}) in (\ref{qtheta_cal}), $\ln q(\theta_i|\mathbf Y)$ is obtained as
\begin{align}\notag
&\ln q(\theta_i|\mathbf Y) = {\rm E}_{q({\mathbf z}\setminus{\theta_i}|{\mathbf Y})}[\ln p({\mathbf Y}, \boldsymbol\Phi)] + {\rm const}\notag\\
=&{\rm E}_{q({\mathbf z}\setminus{\theta_i}|{\mathbf Y})}[\ln(p(\boldsymbol\theta)p(\mathbf s)p(\mathbf {W|s})p(\mathbf {Y}|\boldsymbol \theta,\mathbf W))] + {\rm const}\notag\\
=&{\rm E}_{q({\mathbf z}\setminus{\theta_i}|{\mathbf Y})}[\sum_{j=1}^N \ln p(\theta_j)+\sum_{j=1}^N \ln p(s_j)+\ln p(\mathbf {W|s})+ \ln p(\mathbf {Y}|\boldsymbol \theta,\mathbf W)] + {\rm const}\notag\\
=&\ln p(\theta_i)+{\rm E}_{q({\mathbf z}\setminus{\theta_i}|{\mathbf Y})}[\nu^{-1}{||\mathbf {Y-A_{\widehat{\mathcal S}}}\mathbf W_{\widehat{\mathcal S}}||}_{\rm F}^2]+{\rm const}\notag\\
=&\ln p(\theta_i)+2\nu^{-1}{\rm Re}\left\{{\widehat{\mathbf w}}_i^{\rm T}{\mathbf Y}^{\rm H}\mathbf a(\theta_i)\right\}
-2\nu^{-1}{\rm Re}\left\{{\rm E}_{q({\mathbf z}\setminus{\theta_i}|{\mathbf Y})}\left[({{\mathbf w}^{\rm T}_i}{\mathbf W^{\rm H}_{\widehat{\mathcal S} \backslash \{i\}}}{\mathbf A^{\rm H}_{\widehat{\mathcal S} \backslash \{i\}}})\mathbf a(\theta_i)\right]
\right\}+ {\rm const}\notag\\
\overset{a}=&\ln p(\theta_i)+{\rm Re}\left\{{\boldsymbol \eta}_i^{\rm H}\mathbf a(\theta_i)\right\},
\end{align}
where $\overset{a}=$ utilizes (\ref{w_est}), and the complex vector ${\boldsymbol\eta}_i$ is given in (\ref{yita-i}). Thus $q(\theta_i|\mathbf Y)$ is obtained in (\ref{pdf-q}).
\subsection{Finding a local maximum of $\ln Z({\mathbf s})$}\label{modelSelection}
Finding the globally optimal binary sequence $\mathbf s$ of (\ref{lnZ}) is hard in general. As a result, a greedy iterative search strategy is adopted \cite{Badiu}. We proceed as follows: In the $p$th iteration, we obtain the $k$th test sequence $\mathbf t_k$ by flipping the $k$th element of $\mathbf s^{(p)}$. Then we calculate $\Delta^{(p)}_k=\ln Z(\mathbf t_k)-\ln Z({\mathbf s}^{(p)})$ for each $k=1,\cdots,N$. If $\Delta^{(p)}_k<0$ holds for all $k$ we terminate the algorithm and set $\widehat{\mathbf s}={\mathbf s}^{(p)}$, else we choose the $t_k$ corresponding to the maximum $\Delta^{(p)}_k$ as $\mathbf s^{(p+1)}$ in the next iteration.

When $k\not\in{\mathcal S}$, that is, $s_k=0$, we activate the $k$th component of $\mathbf s$ by setting $s_k^{'}=1$. Now, ${\mathcal S}'=\mathcal S\cup\{k\}$.
\begin{align}\notag\label{delta_z1}
&\Delta_k = \ln Z(\mathbf s')-\ln Z(\mathbf s)\\
=&L\left(\ln\det(\mathbf J_{\mathcal S}+\frac{\nu}{\tau}\mathbf I_{|\mathcal S|})-\ln\det(\mathbf J_{\mathcal S'}+\frac{\nu}{\tau}\mathbf I_{|\mathcal S'|})\right)+\ln\frac{\lambda}{1-\lambda}+L\ln\frac{\nu}{\tau}\notag\\
+&\nu^{-1}{\rm tr}\left(\mathbf H_{\mathcal S'}^{\rm H}(\mathbf J_{\mathcal S'}+\frac{\nu}{\tau}\mathbf I_{|\mathcal S'|})^{-1}\mathbf H_{\mathcal S'}-\mathbf H_{\mathcal S}^{\rm H}(\mathbf J_{\mathcal S}+\frac{\nu}{\tau}\mathbf I_{|\mathcal S|})^{-1}\mathbf H_{\mathcal S}\right).
\end{align}
Let ${\mathbf j}_k = {\mathbf J}_{{\mathcal S},k}$ denote the $k$th column of ${\mathbf J}_{{\mathcal S}}$ and ${\mathbf h}_k^{\rm T} = {\mathbf H}_{k,:}$ denote the $k$th row of $\mathbf H$. Generally, ${\mathbf j}_k$ and ${\mathbf j}_k^{\rm T}$ should be inserted into the $k$th column and $k$th row of ${\mathbf J}_{\mathcal S}$, respectively, and $M$ is inserted into $(k,k)$th of ${\mathbf J}_{\mathcal S}$ to obtain ${\mathbf J}_{{\mathcal S}'}$. By using the block-matrix determinant formula, one has
\begin{align}\label{det_J}
\ln\det(\mathbf J_{\mathcal S'}+\frac{\nu}{\tau}\mathbf I_{|\mathcal S'|})=\ln{\det(\mathbf J_\mathcal S+\frac{\nu}{\tau}\mathbf I_{|\mathcal S|})}+{\rm ln}{\left(M+\frac{\nu}{\tau}-{\mathbf j}_k^{\rm H}(\mathbf J_\mathcal S+\frac{\nu}{\tau}\mathbf I_{|\mathcal S|})^{-1}\mathbf j_k\right)}.
\end{align}
Similarly, ${\mathbf h}_k^{\rm T}$ is inserted into the $k$th row of $\mathbf H_{\mathcal S}$. By the block-wise matrix inversion formula, one has
\begin{align}\label{det_H}
{\rm tr}\left[\mathbf H_{\mathcal S'}^{\rm H}(\mathbf J_{\mathcal S'}+\frac{\nu}{\tau}\mathbf I_{|\mathcal S'|})^{-1}\mathbf H_{\mathcal S'}\right] = {\rm tr}\left[\mathbf H_{\mathcal S}^{\rm H}(\mathbf J_{\mathcal S}+\frac{\nu}{\tau}\mathbf I_{|\mathcal S|})^{-1}\mathbf H_{\mathcal S}\right]+\nu\frac{{\mathbf u}_k^{\rm H}{\mathbf u}_k}{v_k},
\end{align}
where
\begin{align}\label{v-k-u-k}
&v_k = \nu\left(M+\frac{\nu}{\tau}-\mathbf j^{\rm H}_k(\mathbf J_{\mathcal S}+\frac{\nu}{\tau}\mathbf I_{|\mathcal S|})^{-1}{\mathbf j}_k\right)^{-1},\notag\\
&{\mathbf u}_k = \nu^{-1}v_k\left({\mathbf h}_k^*-{\mathbf H}_{\mathcal S}^{\rm H}(\mathbf J_\mathcal S+\frac{\nu}{\tau}\mathbf I_{|\mathcal S|})^{-1}{\mathbf j}_k\right).
\end{align}
Inserting (\ref{det_J}) and (\ref{det_H}) into (\ref{delta_z1}), $\Delta_k$ can be simplified as
\begin{align}\label{delta-k-active}
\Delta_k = L\ln\frac{v_k}{\tau} + \frac{{\mathbf u}^{\rm H}_k{\mathbf u}_k}{v_k}+\ln\frac{\lambda}{1-\lambda}.
\end{align}
Given that $\mathbf s$ is changed into ${\mathbf s}'$, the mean $\widehat{\mathbf W}'_{{\mathcal S'}}$ and covariance ${\widehat{\mathbf C}}_{\mathcal S',0}$ of the weights can be updated from (\ref{W-C-1}), i.e.,
\begin{subequations}
\begin{align}\label{cov_update}
{\widehat{\mathbf C}}_{\mathcal S',0} &= \nu(\mathbf J_{\mathcal S'}+\frac{\nu}{\tau}\mathbf I_{|\mathcal S'|})^{-1},\\
\widehat{\mathbf W}'_{{\mathcal S'}} &= \nu^{-1}\widehat{\mathbf C}'_{{\mathcal S'},0}\mathbf H_{{\mathcal S'}}.
\end{align}
\end{subequations}
In fact, the matrix inversion can be avoided when updating $\widehat{\mathbf W}'_{{\mathcal S'}}$ and ${\widehat{\mathbf C}}_{\mathcal S',0}$. It can be shown that
\begin{align}\label{C_active}
&{\widehat{\mathbf C}}_{\mathcal S',0} =
\begin{pmatrix}
{\widehat{\mathbf C}'_{\mathcal S'\backslash k,0}} & {\widehat{\mathbf c}}'_{k,0} \\
{\widehat{\mathbf c}'^{\rm H}_{k,0}} & \widehat{C}'_{kk,0}
\end{pmatrix}=
\nu\begin{pmatrix}
\mathbf J_\mathcal S+\frac{\nu}{\tau}\mathbf I_{|\mathcal S|} & \mathbf j_k \\
\mathbf j_k^{\rm H} & M+\frac{\nu}{\tau}\\
\end{pmatrix}^{-1}\notag\\
=&
\nu\begin{pmatrix}
 \nu\widehat{\mathbf C}_{\mathcal S,0}^{-1} & \mathbf j_k \\
\mathbf j_k^{\rm H} & M+\frac{\nu}{\tau}\\
\end{pmatrix}^{-1}=
\begin{pmatrix}
{\widehat{\mathbf C}}_{\mathcal S,0}+\frac{v_k}{\nu^2}{\widehat{\mathbf C}}_{\mathcal S,0}\mathbf j_k\mathbf j^{\rm H}_k{\widehat{\mathbf C}}_{\mathcal S,0}  -&\frac{v_k}{\nu}{\widehat{\mathbf C}}_{\mathcal S,0}\mathbf j_k \\
-\frac{v_k}{\nu}\mathbf j^{\rm H}_k{\widehat{\mathbf C}}_{\mathcal S,0} & v_k
\end{pmatrix}.
\end{align}
Furthermore, the weight ${\widehat{\mathbf W}}'_{\mathcal S'}$ is updated as
\begin{align}
&{\widehat{\mathbf W}}'_{\mathcal S'}=\begin{pmatrix}
{\widehat{\mathbf W}'_{\mathcal S'\backslash k}} \\
{\widehat{\mathbf w}'^{\rm T}_k}
\end{pmatrix}=\nu^{-1}
\begin{pmatrix}
{\widehat{\mathbf C}'_{\mathcal S'\backslash k,0}} & {\widehat{\mathbf c}}'_{k,0} \\
{\widehat{\mathbf c}'^{\rm H}_{k,0}} & \widehat{C}'_{kk,0}
\end{pmatrix}
\begin{pmatrix}
{{\mathbf H}_{\mathcal S'\backslash k,0}} \\
{{\mathbf h}^{\rm T}_{k}}
\end{pmatrix} =
\begin{pmatrix}
{\widehat{\mathbf W}}_{\mathcal S} - \nu^{-1}\widehat{\mathbf C}_{\mathcal S,0}{\mathbf j}_k{\mathbf u}^{\rm H}_k\\
{\mathbf u}^{\rm H}_k
\end{pmatrix}.\notag
\end{align}
It can be seen that after activating the $k$th component, the posterior mean and variance of ${\mathbf w}_k$ are ${\mathbf u}_k$ and $v_k{\mathbf I}_L$, respectively.

%

For the deactive case with ${\mathbf s}_k=1$, ${\mathbf s}'_k = 0$ and ${\mathcal S}'=\mathcal S\backslash\{k\}$, $\Delta_k = \ln Z(\mathbf s')-\ln Z(\mathbf s)$ is the negative of (\ref{delta-k-active}), i.e.,
\begin{align}\label{delta-k-deactive}
\Delta_k =- L\ln\frac{v_k}{\tau} - \frac{{\mathbf u}^{\rm H}_k{\mathbf u}_k}{v_k}-\ln\frac{\lambda}{1-\lambda}.
\end{align}
Similar to (\ref{C_active}), the posterior mean and covariance update equation from ${\mathcal S}'$ to ${\mathcal S}$ case can be rewritten as
\begin{align}\label{C_deactive}
\begin{pmatrix}
{\widehat{\mathbf C}'}_{\mathcal S',0}+\frac{v_k}{\nu^2}{\widehat{\mathbf C}'}_{\mathcal S',0}\mathbf j_k\mathbf j^{\rm H}_k{\widehat{\mathbf C}'}_{\mathcal S',0} & -\frac{v_k}{\nu}{\widehat{\mathbf C}'}_{\mathcal S',0}\mathbf j_k \\
-\frac{v_k}{\nu}\mathbf j^{\rm H}_k{\widehat{\mathbf C}'}_{\mathcal S',0} & v_k
\end{pmatrix} =
\begin{pmatrix}
{\widehat{\mathbf C}_{\mathcal S\backslash k,0}} & {\widehat{\mathbf c}}_{k,0} \\
{\widehat{\mathbf c}^{\rm H}_{k,0}} & \widehat{C}_{kk,0}
\end{pmatrix}
\end{align}
\begin{align}\label{W_deactive}
\begin{pmatrix}
{\widehat{\mathbf W}'}_{\mathcal S'} - \nu^{-1}\widehat{\mathbf C}'_{\mathcal S',0}{\mathbf j}_k{\mathbf u}^{\rm H}_k\\
{\mathbf u}^{\rm H}_k
\end{pmatrix}=
\begin{pmatrix}
{\widehat{\mathbf W}_{\mathcal S\backslash k}} \\
{\widehat{\mathbf w}^{\rm T}_k}
\end{pmatrix},
\end{align}
where $\widehat{\mathbf c}_{k,0}$ denotes the column of ${\widehat{\mathbf C}}_{\mathcal S,0}$ corresponding to the $k$th component. According to (\ref{C_deactive}) and (\ref{W_deactive}), one has
\begin{subequations}\label{C_de_appro}
\begin{align}
{\widehat{\mathbf C}'}_{\mathcal S',0}+\frac{v_k}{\nu^2}{\widehat{\mathbf C}'}_{\mathcal S',0}\mathbf j_k\mathbf j^{\rm H}_k{\widehat{\mathbf C}'}_{\mathcal S',0}&={\widehat{\mathbf C}_{\mathcal S\backslash k,0}},\label{C_de_approa}\\
-\frac{v_k}{\nu}{\widehat{\mathbf C}'}_{\mathcal S',0}{\mathbf j}_k&= {\widehat{\mathbf c}}_{k,0}\label{C_de_approb}\\
v_k &= \widehat{C}_{kk,0},\label{C_de_approc}\\
{\widehat{\mathbf W}'}_{\mathcal S'} - \nu^{-1}\widehat{\mathbf C}'_{\mathcal S',0}{\mathbf j}_k{\mathbf u}^{\rm H}_k&={\widehat{\mathbf W}_{\mathcal S\backslash k}},\label{C_de_approd}\\
{\mathbf u}^{\rm H}_k&={\widehat{\mathbf w}^{\rm T}_k}.\label{C_de_approe}
\end{align}
\end{subequations}
Thus, ${\widehat{\mathbf C}'}_{\mathcal S',0}$ can be updated by substituting (\ref{C_de_approb}) and (\ref{C_de_approc}) in (\ref{C_de_approa}), i.e.,
\begin{align}
{\widehat{\mathbf C}'}_{\mathcal S',0} = {\widehat{\mathbf C}_{\mathcal S\backslash k,0}} - \frac{v_k}{\nu^2}{\widehat{\mathbf C}'}_{\mathcal S',0}\mathbf j_k\mathbf j^{\rm H}_k{\widehat{\mathbf C}'}_{\mathcal S',0} = {\widehat{\mathbf C}_{\mathcal S\backslash k,0}} - \frac{{\widehat{\mathbf c}}_{k,0}{\widehat{\mathbf c}}^{\rm H}_{k,0}}{\widehat{C}_{kk,0}}.
\end{align}
Similarly, ${\widehat{\mathbf W}'}_{\mathcal S'}$ can be updated by substituting (\ref{C_de_approb}) and (\ref{C_de_approe}) in (\ref{C_de_approd}), i.e.,
\begin{align}
{\widehat{\mathbf W}'}_{\mathcal S'} = \nu^{-1}\widehat{\mathbf C}'_{\mathcal S',0}{\mathbf j}_k{\mathbf u}^{\rm H}_k + {\widehat{\mathbf W}_{\mathcal S\backslash k}} = {\widehat{\mathbf W}_{\mathcal S\backslash k}} -  \frac{{\widehat{\mathbf c}}_{k,0}}{{\widehat{C}}_{kk,0}}\widehat{\mathbf w}^{\rm T}_k.
\end{align}
According to $v_k = \widehat{C}_{kk,0}$ (\ref{C_de_approc}) and ${\mathbf u}^{\rm H}_k = {\widehat{\mathbf w}^{\rm T}_k}$ (\ref{C_de_approe}), $\Delta_k$ (\ref{delta-k-deactive}) can be simplified as
\begin{align}\label{delta-k-deactivesim}
\Delta_k =-L\ln\frac{{\widehat{C}}_{kk,0}}{\tau} - \frac{{\mathbf w}^{\rm H}_k{\mathbf w}_k}{{\widehat{C}}_{kk,0}}-\ln\frac{\lambda}{1-\lambda}.
\end{align}
\subsection{Estimation of model parameters}\label{estmopa}
Plugging the postulated PDF (\ref{postpdf}) in (\ref{DL-L}), one has
\begin{align}
&{\mathcal L}(q{(\boldsymbol\theta\mathbf {,W,s|Y})};\boldsymbol \beta)= {\rm E}_{q{(\boldsymbol\theta\mathbf {,W,s|Y}})}\left[\ln{\tfrac{p({\mathbf {Y,}\boldsymbol\theta\mathbf{,W,s};\boldsymbol \beta})}{q({\boldsymbol\theta\mathbf {,W,s|Y}})}}\right]\notag\\
=& {\rm E}_{q{(\boldsymbol\theta\mathbf {,W,S|Y}})}\left[\sum_{i=1}^N \ln p(s_i) + \ln p(\mathbf {W|s})+\ln p(\mathbf {Y}|\boldsymbol \theta,\mathbf W)\right]+{\rm const}\notag\\
=&||{\widehat{\mathbf s}||_0\ln\lambda} -||{\widehat{\mathbf s}||_0\ln(1-\lambda)} + ||\widehat{\mathbf s}||_0L\ln\frac{1}{\pi\tau}
-{\rm E}_{q({\mathbf {W|Y}})}\left[\frac{1}{\tau}{\rm tr}(\mathbf W_{\widehat{\mathcal S}}\mathbf W^{\rm H}_{\widehat{\mathcal S}})\right]+ {\rm const} \notag\\
+&ML\ln\frac{1}{\pi\nu}-\frac{1}{\nu}{\rm tr}(\mathbf Y^{\rm H}\mathbf Y)+\frac{2}{\nu}{\rm Re}\{{\rm tr}(\widehat{\mathbf W}^{\rm H}_{\widehat{\mathcal S}}\mathbf H_{\widehat{\mathcal S}})\}-\frac{1}{\nu}{\rm E}_{q{(\mathbf {W|Y}})}[{\rm tr}(\mathbf W^{\rm H}_{\widehat{\mathcal S}}\mathbf J_{\widehat{\mathcal S}}\mathbf W_{\widehat{\mathcal S}})].\notag
\end{align}
Substituting ${\rm E}_{q{(\mathbf {W|Y}})}[{\rm tr}(\mathbf W_{\widehat{\mathcal S}}\mathbf W^{\rm H}_{\widehat{\mathcal S}})]={\rm tr}(\widehat{\mathbf W}^{\rm H}_{\widehat{\mathcal S}}\widehat{\mathbf W}_{\widehat{\mathcal S}})+L{\rm tr}({\widehat{\mathbf C}}_{\widehat{\mathcal S},0})$ and ${\rm E}_{q{(\mathbf {W|Y}})}[{\rm tr}(\mathbf W^{\rm H}_{\widehat{\mathcal S}}\mathbf J_{\widehat{\mathcal S}}\mathbf W_{\widehat{\mathcal S}})]={\rm tr}(\mathbf J_{\widehat{\mathcal S}}(\widehat{\mathbf W}_{\widehat{\mathcal S}}\widehat{\mathbf W}^{\rm H}_{\widehat{\mathcal S}}+L\widehat{{\mathbf C}}_{\widehat{\mathcal S},0}))$ in the above equation, ${\mathcal L}(q{(\boldsymbol\theta\mathbf {,W,s|Y})};\boldsymbol \beta)$ is obtained as (\ref{Lvaluemodpa}).
\section{Acknowledgement}
The authors thank Qian Cheng for sharing the EPUMA code to help us make the performance comparison in the DOA experiments.

\bibliographystyle{IEEEbib}
\bibliography{strings,refs}

\begin{thebibliography}{}
\bibitem{Confv}
Q. Zhang, J. Zhu, P. Gerstoft, M. A. Badiu and Z. Xu, ``Gridless Line Spectral Estimation with Multiple Measurement Vector via Variational Bayesian Inference,'' submitted to ICASSP 2019.
\bibitem{Stoica}
P. Stoica and R. L. Moses, \emph{Spectral Analysis of Signals.} Upper Saddle River, NJ, USA: Prentice-Hall, 2005.
\bibitem{Bajwa}
W. Bajwa, A. Sayeed, and R. Nowak, ``Compressed channel sensing:
A new approach to estimating sparse multipath channels,'' \emph{Proc. IEEE}, vol. 98, pp. 1058-1076, Jun. 2010.
\bibitem{HansenCom}
T. L. Hansen, P. B. J{\o}rgensen, M. A. Badiu and B. H. Fleury, ``An iterative receiver for OFDM with sparsity-based parametric channel estimation,'' \emph{IEEE Trans. Signal Process.}, vol. 66, no. 20, pp. 5454-5469, 2018.
\bibitem{Ottersten}
B. Ottersten, M. Viberg and T. Kailath, ``Analysis of subspace fitting
and ML techniques for parameter estimation from sensor array data,'' \emph{IEEE Trans. Signal Process.}, vol. 40, pp. 590-600, Mar. 1992.
\bibitem{Schmidt}
R. Schmidt, ``Multiple emitter location and signal parameter estimation,'' \emph{IEEE Trans. on Antennas and Propagation}, vol. 34, no. 3, pp. 276-280, 1986.
\bibitem{Roy}
R. Roy and T. Kailath, ``ESPRIT - estimation of signal parameters via rotational invariance techniques,''  \emph{IEEE Trans. on Acoustics, Speech and Signal Processing}, vol. 37, no. 7, pp. 984-995, 1989.
\bibitem{StociaML}
P. Stoica and A. Nehorai, ``Music, maximum likelihood and Cram\'{e}r-Rao bound: further results and comparisons,'' \emph{IEEE Trans. Acoust., Speech, Signal Processing}, vol. 38, no. 12, pp. 2140-2150, Dec. 1990.
\bibitem{Wax1}
I. Ziskind and M. Wax, ``Maximum likelihood localization of multiple
sources by alternating projection,'' \emph{IEEE Trans. Acoust., Speech, Signal
Process.}, vol. 36, no. 10, pp. 1553¨C1560, Oct. 1988.
\bibitem{Fleury1}
B.H. Fleury, M. Tschudin, R. Heddergott, D. Dahlhaus, and K. I. Pedersen,
``Channel parameter estimation in mobile radio environments using the
SAGE algorithm,'' \emph{IEEE J. Sel. Areas Commun.}, vol. 17, no. 3, pp. 434-450, Mar. 1999.
\bibitem{Stoica1}
P. Stoica and Y. Selen, ``Model-order selection: a review of information criterion rules,'' \emph{IEEE Signal Processing Magazine}, vol. 21, no. 4, pp. 36-47, July 2004.
\bibitem{Malioutov}
D. Malioutov, M. Cetin and A. Willsky, ``A sparse signal reconstruction perspective for source localization with sensor arrays,'' \emph{IEEE Trans. Signal Process.}, vol. 53, no. 8, pp. 3010-2022, 2005.
\bibitem{Stoica2}
P. Stoica, P. Babu and J. Li, ``New method of sparse parameter estimation in separable models and its use for spectral analysis of irregularly sampled data,'' \emph{IEEE Trans. Signal Process.}, vol. 59, no. 1, pp. 35-47, Jan. 2011.
\bibitem{Stoica3}
P. Stoica, P. Babu and J. Li, ``SPICE: A sparse covariance-based estimation method for array processing,'' \emph{IEEE Trans. Signal Process.}, vol. 59, no. 2, pp. 629-638, Feb. 2011.
\bibitem{Stoica4}
P. Stoica and P. Babu, ``SPICE and LIKES: Two hyperparameter-free methods for sparse-parameter estimation,'' \emph{Signal Process.}, vol. 92, no. 7, pp. 1580-1590, 2012.
\bibitem{Gerstoft}
P. Gerstoft, C. F. Mecklenbr\"{a}uker, A. Xenaki, and S. Nannuru, ``Multisnapshot sparse Bayesian learning for DOA,'' \emph{IEEE Signal Process. Lett.}, vol 23, no. 10, pp. 1469-1473, 2016.
\bibitem{Hu}
L. Hu, Z. Shi, J. Zhou and Q. Fu, ``Compressed sensing of complex sinusoids: An approach based on dictionary refinement'' \emph{IEEE Trans. Signal Process.}, vol. 60, no. 7, pp. 3809-3822, 2012.
\bibitem{Mamandipoor}
B. Mamandipoor, D. Ramasamy and U. Madhow, ``Newtonized orthogonal matching pursuit: Frequency estimation over the continuum,'' \emph{IEEE Trans. Signal Process.}, vol. 64, no. 19, pp. 5066-5081, 2016.
\bibitem{Lin}
J. Zhu, L. Han, R. S. Blum and Z. Xu, ``Newtonized orthogonal matching pursuit for line spectrum estimation with multiple measurement vectors,'' avaliable at https://arxiv.org/pdf/1802.01266.pdf.
\bibitem{Fang}
J. Fang, F. Wang, Y. Shen, H. Li and R. S. Blum, ``Superresolution compressed sensing for line spectral estimation:an iterative reweighted approach,'' \emph{IEEE Trans. Signal Process.}, vol. 64, no. 18, pp. 4649-4662, 2016.
\bibitem{Tang}
G. Tang, B. Bhaskar, P. Shah and B. Recht, ``Compressed sensing off the grid,'' \emph{IEEE Trans. Inf. Theory}, vol. 59, no. 11, pp. 7465-7490, 2013.
\bibitem{Bhaskar}
B. Bhaskar, G. Tang, and B. Recht, ``Atomic norm denoising with applications to line spectral estimation,'' \emph{IEEE Trans. Signal Process.}, vol. 61,
no. 23, pp. 5987-5999, Dec. 2013.
\bibitem{Yang2}
Z. Yang and L. Xie, ``On gridless sparse methods for line spectral estimation from complete and incomplete data,'' \emph{IEEE Trans. Signal Process.},  vol. 63, no. 12, pp. 3139-3153, 2015.
\bibitem{Yang1}
Z. Yang, L. Xie and C. Zhang, ``A discretization-free sparse and parametric approach for linear array signal processing,''  \emph{IEEE Trans. Signal Process.}, vol. 62, no. 19, pp. 4959-4973, 2014.
\bibitem{Yang3}
Z. Yang and L. Xie, ``Continuous compressed sensing with a single or multiple measurement vectors,'' \emph{IEEE Workshop on Statistical Signal Processing}, pp. 288-291, 2014.
\bibitem{Chi}
Y. Li and Y. Chi, ``Off-the-grid line spectrum denoising and estimation with multiple measurement vectors,'' \emph{IEEE Trans. Signal Process.}, vol. 64, no. 5, pp. 1257-1269, 2016.
\bibitem{EMac}
Y. Chen and Y. Chi, ``Robust spectral compressed sensing via structured matrix completion,'' \emph{IEEE Trans. Inf. Theory}, vol. 60, no. 10, pp. 6576-6601, Oct. 2014.
\bibitem{RAM}
Z. Yang and L. Xie, ``Enhancing sparsity and resolution via reweighted
atomic norm minimization,'' \emph{IEEE Trans. Signal Process.}, vol. 64, no. 4,
pp. 995-1006, Feb. 2016.
\bibitem{boyd}
S. Boyd and L. Vandenberghe, \emph{Convex Optimization}, Cambridge University Press, 2004.
\bibitem{Tipping}
M. E. Tipping, ``Sparse Bayesian learning and the relevance vector machine,'' \emph{J. Mach. Learn. Res.}, vol. 1, pp. 211-244, 2001.
\bibitem{Wipf}
D. P. Wipf and B. D. Rao, ``Sparse Bayesian learning for basis selection,''
\emph{IEEE Trans. Signal Process.}, vol. 52, no. 8, pp. 2153-2164, Aug. 2004.
\bibitem{Shution}
D. Shutin and B. H. Fleury, ``Sparse variational Bayesian SAGE algorithm with application to the estimation of multipath wireless channels,'' \emph{IEEE
Trans. Signal Process.}, vol. 59, no. 8, pp. 3609-3623, Aug. 2011.
\bibitem{Hansenconf}
T. L. Hansen, M. A. Badiu, B. H. Fleury, and B. D. Rao, ``A sparse
Bayesian learning algorithm with dictionary parameter estimation,'' \emph{in
Proc. IEEE 8th Sensor Array Multichannel Signal Process. Workshop},
Jun. 2014, pp. 385-388.
\bibitem{Badiu}
M. A. Badiu, T. L. Hansen and B. H. Fleury, ``Variational Bayesian inference of line spectral,'' \emph{IEEE Trans. Signal Process.}, vol. 65, no. 9, pp. 2247-2261, 2017.
\bibitem{HansenTSP}
T. L. Hansen, B. H. Fleury and B. D. Rao, ``Superfast line spectral estimation,'' avaliable at https://arxiv.org/pdf/1705.06073.pdf.
\bibitem{Dave}
D. Zachariah, P. Wirf\"{a}lt, M. Jansson and S. Chatterjee, ``Line spectrum estimation with probabilistic priors,'' \emph{Signal Processing}, vol. 93, no. 11, pp. 2969-2974, 2013.
\bibitem{Direc}
K. V. Mardia and P. E. Jupp, \emph{Directional Statistics}. New York, NY, USA:
Wiley, 2000.
\bibitem{Murphy}
K. P. Murphy, \emph{Machine Learning A Probabilistic Perspective}. MIT Press, 2012.
\bibitem{Bertsekas}
D. P. Bertsekas and J. N. Tsitsiklis : \emph{Parallel and Distributed Computation: Numerical Methods}, Athenan Scientific: Massachusetts, 1997.
\bibitem{Are}
A. Hj$\phi$rungnes, \emph{Complex-Valued Matrix Derivatives: With Applications in Signal Processing and Communications}, Cambridge University Press, 2011.
\bibitem{Huang}
C. Qian, L. Huang, N. D. Sidiropoilos and H. C. So, ``Enhanced PUMA for direction-of-arrival estimation and its performance analysis,'' \emph{IEEE Trans. Signal Process.}, vol. 64, no. 16, pp. 4127-4137, 2016.
\bibitem{SeqDOA}
C. F. Mecklenbrauker, P. Gerstoft, A. Panahi, and M. Viberg, ``Sequential
Bayesian sparse signal reconstruction using array data,'' \emph{IEEE Trans.
Signal Process.}, vol. 61, no. 24, pp. 6344-6354, 2013.
\bibitem{Meng}
X. Meng, S. Wu and J. Zhu, ``A unified Bayesian
inference framework for generalized linear model,'' \emph{IEEE Signal Process. Lett.}, vol. 25, no. 3, Mar. 2018.
\bibitem{offVALSEq}
J. Zhu, Q. Zhang and X. Meng, ``Off-grid variational Bayesian inference of line spectral estimation from one-bit samples,'' avaliable at https://arxiv.org/pdf/1811.05680.pdf.
\end{thebibliography}

\end{document}